\documentclass[journal=jacsat,manuscript=article]{achemso}
\usepackage[T1]{fontenc} % Use modern font encodings
\usepackage[dvipsnames]{xcolor}
\usepackage{amsmath, mathtools}
\usepackage{amsfonts}
\usepackage{braket}
\usepackage{siunitx}
\usepackage{graphicx}
\usepackage{multicol}
\usepackage{booktabs}
\usepackage{tikz}
\usepackage{subfig}
\usepackage{tensind}
\tensorformat{lrb}
\tensordelimiter{?}
\usepackage{pgfplots}
\pgfplotsset{every axis/.append style={very thick,  tick style={ultra thick}}, compat=1.9, width=8.7cm}
\usetikzlibrary{pgfplots.groupplots}
\usepackage{threeparttable}
\usepackage{achemso}
\mciteErrorOnUnknownfalse

\usetikzlibrary{matrix,positioning,decorations.pathreplacing}
\newenvironment{Figure}
  {\par\medskip\noindent\minipage{\linewidth}}
  {\endminipage\par\medskip}

\author{Nicholas Lee}
\author{Alex J. W. Thom}
\email{ajwt3@cam.ac.uk}
\affiliation[University of Cambridge]
{Yusuf Hamied Department of Chemistry, Lensfield Road, Cambridge, CB2 1EW, U.K.}

\title {Localised Spin Rotations: A Size-Consistent Approach to Non-Orthogonal Configuration Interaction}

\begin{document}
\maketitle

\begin{abstract}
Current Non-Orthogonal Configuration Interaction (NOCI) methods often use a set of Self-Consistent Field (SCF) states selected based on chemical intuition.  However, it may be challenging to track these SCF states across a dissociation profile and the NOCI states recovered may be spin contaminated. In this paper, we propose a method of applying spin rotation on symmetry broken UHF (sb-UHF) states to generate a basis for NOCI. The dissociation of ethene was examined by localising spin rotation on each resulting carbene fragment. We show that this gives a size-consistent description of its dissociation and results in spin-pure states at all geometries. The dissociation was also studied with different orbitals, namely canonical UHF and Absolutely Localised Molecular Orbitals (ALMO). Furthermore, we demonstrate that the method can be used to restore spin symmetry of symmetry broken SCF wavefunctions for molecules of various sizes, marking an improvement over existing NOCI methods.
\end{abstract}

\section{Introduction}
Hartree-Fock theory is a mean-field approach where each electron interacts with the averaged field generated by all other electrons\cite{Szabo}. However, the mean-field approach is an approximation and need not, in general, respect the symmetries of the electronic Hamiltonian\cite{Bang}. As such, the self-consistent solutions found using the Hartree-Fock equations may not respect these same symmetries, leading to symmetry breaking\cite{IPM1, Russ, IPM2}. A classic example comes from Unrestricted Hartree-Fock (UHF) where the $\alpha$ and $\beta$ spin orbitals are allowed to be described with different spatial functions. This typically leads to a symmetry broken UHF (sb-UHF) solution\cite{Amos-Hill} at a lower energy level than the Restricted Hartree-Fock (RHF) solution.\\
When modelling not just a simple molecule, but a dissociating one, it is well-known\cite{Helgaker} that the Restricted Hartree-Fock (RHF) description of a molecular system often breaks down as bonds dissociate. The Unrestricted Hartree-Fock (UHF) description gives a more accurate representation of the dissociation, but the UHF solutions are not in general eigenfunctions of $\hat{S}^{2}$. While the UHF solution gives a lower energy and is thus a closer approximation to the true ground state wave function by the variational principle, it comes at the cost of spin contamination\cite{SPUHF, SpinContamination}. 
Both the examples of a simple molecular system and a dissociating one are a manifestations of what L\"{o}wdin referred to as the \emph{Symmetry Dilemma}\cite{SymDilemma}. \\
Spin contamination is also a serious concern when attempting to assign spin states in a molecular system. For example, in singlet fission research\cite{Smith1, Smith2, Casanova, Havenith1, Zimmerman}, the singlet-triplet gap is of paramount importance as it determines the thermodynamics, and hence feasibility of the fission. An accurate prediction of the singlet-triplet gap is current being sought after in the field, and so there is a great desire to obtain states with a well-defined total spin.\\
Ideally, we seek a method that gives spin states with good quantum numbers, is size extensive, and scales well with system size. To that end, there have been quite a number of approaches developed. Projected Hartree-Fock (PHF)\cite{PQT, PHF} and Half-Projection methods\cite{HPT, HPTGeminals} were developed to restore various symmetries such as spin and particle numbers.
These are known as Variation after Projection (VAP) approaches.
Another promising method for producing spin eigenstates is Neuscamman's Jastrow-modified Antisymmetric Geminal Power (JAGP)\cite{JAGP}, and brings on other desirable properties such as size-extensivity and polynomial scaling. However, the implementation of this procedure is challenging. \\
In contrast we propose a method where the symmetry is restored after a variational approach, so share the philosophy of Projection after Variation (PAV).  However, using NOCI, the symmetry restoration projection is performed variationally using a Configuration Interaction, so we denote our methods Variational Projection after Variation (VPAV).
%Modified JAGP paragraph
Recently, our group has reported the use of Hartree-Fock (HF) solutions found by SCF metadynamics\cite{Thom1, Thom2} as a basis for Non-Orthogonal Configuration Interaction\cite{Kris, Hugh-NOCI}. We refer to such an approach to NOCI as multi-state NOCI (MS-NOCI) to reflect its use of multiple SCF states. HF solutions were used as they are more chemically relevant to the system in question. However, because it is based on the sb-UHF states, this approach may not produce spin states with correct spin symmetry even after a judicious selection of HF solutions. With an aim to generate spin symmetry-respecting states in a general fashion, we demonstrate that using multiple spin-rotations of a single HF state as a basis for NOCI, hereafter known as spin-rotated NOCI (SR-NOCI), is a general and size-extensive method for spin restoration of symmetry broken HF states.\\
It should be noted that the use of spin rotation for spin projection is not new\cite{Percus-Rotenberg, Lefebvre-Prat, Bendazzoli}. Furthermore, the use of spin-rotated states as a basis for NOCI has previously been attempted by Sundstrom and Head-Gordon\cite{Sundstrom-NOCI}, and more recently by Nite and Jimenez-Hoyos\cite{Nite}.  %\AJWT{There are probably older references to things too.}
 This work aims to generalise the approach to dissociating systems and differs from the latter publication in that symmetry-projection is performed through NOCI, without the need for grid-integration. To the best of our knowledge, a comprehensive study on the feasibility of spin-rotated states as a basis for NOCI has not previously been attempted. \\
The remainder of the paper is structured as follows: Section 2 introduces the theory behind the approach and shows that this approach is formally size-consistent. Section 3 considers the method applied to the hydrogen molecule and its dimer, while Section 4 studies the dissociation of ethene. Section 5 compares the results from MS-NOCI against SR-NOCI. We conclude with a summary of the applicability of this method. 

\section{Theory}
	\subsection{General approach}
\begin{tabular}[H]{p{0.2 \textwidth} p{0.7\textwidth}}
\toprule \toprule
Notation & Description \\
\hline
$\chi_{\mu}$ & Atomic orbital $\mu$, where $ 1 \leq \mu \leq M$ for molecule A and $ M+1 \leq \mu \leq M+N$ for molecule B. \\
$U$, $V$ & Single determinant wave functions of complete system of two coupled molecules, A and B.\\
$U^{A}$, $U^{B}$, $V^{A}$, $V^{B}$ & Single determinant wave functions of the respective molecules A and B, in states U and V. \\
$u_{i}$, $v_{i}$ & Spin orbitals belonging to the states U and V, respectively. $ 1 \leq i \leq P$ for molecule A and $ P+1 \leq i \leq P+Q$ for molecule B.  \\
$?[l]c^k_i?$, $?[l]d^l_j?$ & Expansion coefficients for spin orbitals $u_{i}$ and $v_{j}$, respectively, in terms of basis orbitals $\chi_{k}$ and $\chi_{l}$. \\
$\boldsymbol{S}_{UV}^{A}$, $\boldsymbol{S}_{UV}^{B}$ & Overlap matrices between states U and V on molecule A and on molecule B, respectively. \\
$D( i | j )$ &Cofactor of determinant $D$ with row $i$ and column $j$ removed.\\
$\hat{F}$ & One-electron component of the Hamiltonian. \\
$\hat{G}$ & Two-electron component of the Hamiltonian. \\
$\hat{R}(\theta \boldsymbol{\hat{n}})$ & Spin rotation operator of angle $\theta$ about axis $\boldsymbol{\hat{n}}$. \\
$\boldsymbol{S}_{AB}$ & Overlap matrix of the complete system of molecules A and B. \\
$\boldsymbol{S}_{A}$, $\boldsymbol{S}_{B}$ & Overlap matrix of molecule A and B, respectively. \\
$\boldsymbol{H}_{AB}$ & Hamiltonian matrix of the complete system of molecules A and B. \\
$\boldsymbol{H}_{A}$, $\boldsymbol{H}_{B}$ & Hamiltonian matrix of molecule A and B, respectively. \\
\bottomrule \bottomrule
\end{tabular}
\captionof{table}{Notation used. Most of the notation used follows I. Mayer\cite{Mayer}}
\label{tab:Notation} 
To restore the required spin-symmetry, we begin with a sb-UHF state, and perform multiple spin rotations on it to obtain a set of spin-rotated states. The spin-rotated states will thus serve as the basis for NOCI. In the case of a dimer, we perform spin-rotations on the atomic orbitals belonging to each of its constituent molecules independently of each other.
If sufficient rotations are chosen, the set of spin-rotated states is expected to span the spin space sufficiently such that a linear combination of these states will have good spin quantum numbers.\\

In the single molecule case, this is equivalent to the PHF method, where we can draw a parallel between the grid integration done in PHF and spin rotations performed in SR-NOCI. However, while PHF optimises the orbital coefficients after spin-projection (a VAP approach), we use NOCI find the spin-projection coefficients (a VPAV approach).\\
In the case of a dimer, the ability to perform spin-rotations independently on the atomic orbitals centred on each constituent molecule allows for size-consistency. This therefore addresses the problem of size-inconsistency with the PHF method\cite{PHF}.
The mathematical details of this approach are elaborated upon in this section. The notation used for the rest of the paper is summarised in Table \ref{tab:Notation}. Where necessary, we employ Head-Gordon \emph{et al.}'s tensor notation\cite{HG-Tensor}.

	\subsection{Generator Coordinate Method (GCM)}
The GCM has been widely used in nuclear physics since the 1930s to restore symmetry to symmetry-broken wave functions\cite{Peierls-Thouless, GCMwf, GCMmicroscopic, Ring}. Recently, work by Scuseria and co-workers used the method in Projected Hartree-Fock theory\cite{PHF}. As such, a derivation of generalised eigenvalue equation from the ansatz of GCM shall not be included. We shall only point out the important aspects of the derivation. 
%\AJWT{You need to be more explicit about what the `matrices' indices are --- these look like just numbers to me.}
For arbitrary spin state $P$, let state $Q$ be a spin-rotation of it, $\ket{Q} = \hat{R}(\theta \boldsymbol{\hat{n}})\ket{P}$, such that state $Q$ is related to state $P$ by a spin rotation $\theta$ about axis $\boldsymbol{\hat{n}}$. The matrix elements of the Hamiltonian and Overlap between $P$ and $Q$ can then be expressed as:
\begin{equation}
\begin{split}
	H_{pq} &= \bra{P} \hat{H}\ket{Q} \\
	&= \bra{P} \hat{H} \hat{R}(\theta \boldsymbol{\hat{n}}) \ket{P}
\end{split}
\end{equation}
\begin{equation}
\begin{split}
	S_{pq} &= \braket{P | Q} \\
	&= \bra{P} \hat{R}(\theta, \boldsymbol{\hat{n}})\ket{P}
\end{split}
\end{equation}
and define square matrices $\boldsymbol{H}$ and $\boldsymbol{S}$ respectively.
We may create a variationally optimized combination of these states by solving the resulting generalised eigenvalue equation (also known as the discretised Hill-Wheeler equation):
\begin{equation}
		\sum_{q} H_{pq} D_{q i} =  \sum_{q} S_{p q} D_{q i} E_{i}
\label{NOCI}
\end{equation}
for each reference state $P$.
This corresponds to the NOCI equation with spin rotated states of $P$ as the non-orthogonal basis, giving us an early indication that NOCI serves as a discretised version of the GCM and hence a viable candidate for restoring spin symmetry.

	\subsection{Generating Spin-Rotated States}
For a rotation about the axis ${\boldsymbol{\hat n}}=(n_{x}, n_{y}, n_{z})$ by an angle $\theta$, we define the spin rotation matrix $\boldsymbol{R}$ in the basis of $(\alpha, \beta)$ as:
\begin{equation}
\begin{split}
	\boldsymbol{R} &= \begin{pmatrix}
								 		\cos\frac{\theta}{2} - in_{z}\sin\frac{\theta}{2} & (-n_{y} - in_{x})\sin\frac{\theta}{2} \\
								 		(n_{y} - in_{x})\sin\frac{\theta}{2}  & \cos\frac{\theta}{2} + in_{z}\sin\frac{\theta}{2}
							   \end{pmatrix} \\
						   &= \begin{pmatrix}
								 		R_{11} & R_{12} \\
								 		R_{21} & R_{22}
							   \end{pmatrix} \\
\end{split}
\label{RotationMatrix}
\end{equation}
This matrix performs spin rotation on each spin orbital of a state. This is one of the many equivalent forms that have been used in literature where spin-projection is concerned. We give a simple example of one of forms in the Appendix. A comprehensive review on rotation operators is given by Morrison and Parker\cite{Morrison}. For a multi-electron state $P$ and its spin-rotated form $Q = \boldsymbol{R} P$, we can express them in spinor form (The basis is expressed as the direct product of the spatial basis and the two-component spin basis) as follows\cite{Tannoudji}:
\begin{equation}
\begin{split}
\begin{pmatrix}
	Q_{\alpha} \\
	Q_{\beta}
\end{pmatrix} 
&= \boldsymbol{R}
\begin{pmatrix}
	P_{\alpha} \\
	P_{\beta}
\end{pmatrix} \\
&= \begin{pmatrix}
	   R_{11}P_{\alpha} + R_{12}P_{\beta}\\
	   R_{21}P_{\alpha} + R_{22}P_{\beta}
    \end{pmatrix} \\
\end{split}
\end{equation}
where we have partitioned the (Generalized Hartree--Fock (GHF)) orbitals of states $P$ and $Q$ into their spin components.
 $P_{\alpha}$ and $P_{\beta}$ are both matrices of dimensions $ n_{AO} \times n_{e}$ such that $n_{AO}$ is the number of atomic orbitals in the given basis and $n_{e}$ is the total number of electrons.
  For an initial UHF state, for example, $P_\alpha$ consists of a block corresponding to the $\alpha$ orbitals, and contains zeroes for the $\beta$ orbitals.
$Q_{\alpha}$ and $Q_{\beta}$ are similarly matrices of dimensions $ n_{AO} \times n_{e}$, corresponding to the alpha spin and beta spin components respectively. By stacking $Q_\alpha$ and $Q_\beta$ together as a $2n_{AO} \times n_{e}$ matrix, each column corresponds to a GHF orbital. The resultant matrix is then orthogonalised.\\
The expression in (5) implies that the rotation matrix $\boldsymbol{R}$ will have no effect on a RHF coefficient matrix. This is because the alpha and beta coefficients are identical in RHF. Therefore, applying the rotation matrix and orthogonalising has no net effect on the RHF coefficient matrix. The proposed method therefore does not affect RHF solutions.

	\subsection{Proof of Size-Extensivity}
	We consider the case where molecules A and B are far apart, such that $\braket { \chi_{\mu} | \chi_{\nu} } = 0$ if:
\begin{enumerate}
	\item $1 \leq \mu \leq M$ and $ M+1 \leq \nu\leq M+N$, or
	\item $1 \leq \nu \leq M$ and $ M+1 \leq \mu \leq M+N$
\end{enumerate}
i.e. if the two basis orbitals are localized on different molecules.
\subsubsection{Inner product of spin orbitals}
We express the spin orbitals which make up states $U$ and $V$ as a linear combination of basis orbitals.
\begin{equation}
	u_{i} = \sum_{\mu}^{M+N} \chi_{\mu} ?[l]c^{\mu}_i?
\end{equation}
\begin{equation}
	v_{j} = \sum_{\nu}^{M+N} \chi_{\nu} ?[l]d^{\nu}_j?
\end{equation}
The spin orbitals are expressed in the combined basis of molecules A and B, but are localized so they describe either molecule A or B, and the properties of the terms in the equation are summarised in Table \ref{table: Vanishing coefficients}.
\begin{table*}
\centering
\begin{tabular}{ c c c }
 \hline \hline
Molecule & $u_{i} $ &  $v_{j}$ \\
\hline
A  ($ 1 \leq i, j \leq P$)  &   $?[l]c^{\mu}_i? = 0$ for $\mu \geq M + 1$  &  $?[l]d^{\nu}_j? = 0$ for $\nu \geq M + 1$\\

B  ($ P+1 \leq i, j \leq P+Q$)  &   $?[l]c^{\mu}_i? = 0$ for $\mu \leq M$  &  $?[l]d^{\nu}_j?= 0$ for $\nu \leq M$   \\
\hline \hline
\end{tabular}
\caption{Summary of vanishing coefficients.}
\label{table: Vanishing coefficients}
\end{table*}

In this basis of orbitals describing \textbf{both} molecules A and B, the coefficients of the orbitals describing B will be zero on the spin orbitals centered on A and vice versa. \\
Taking the inner product of the two spin orbitals,
\begin{equation*}
\begin{split}
	\braket { u_{i} | v_{j} } &= \sum_{\mu=1}^{M+N} \sum_{\nu=1}^{M+N} ?[r]c*_i^\mu? \braket { \chi_{\mu} | \chi_{\nu} } ?[l]d^{\nu}_j? \\
	&= \!\begin{multlined}[t] 
	\sum_{\mu=1}^{M} \sum_{\nu=1}^{M} ?[r]c*_i^\mu? \braket { \chi_{\mu} | \chi_{\nu} } ?[l]d^{\nu}_j? \\ 
	 +\sum_{\mu=M+1}^{M+N} \sum_{\nu=1}^{M} ?[r]c*_i^\mu?\braket { \chi_{\mu} | \chi_{\nu} } ?[l]d^{\nu}_j? \\
	 + \sum_{\mu=1}^{M} \sum_{\nu=M+1}^{M+N} ?[r]c*_i^\mu? \braket { \chi_{\mu} | \chi_{\nu} } ?[l]d^{\nu}_j? \\ 
	 + \sum_{\mu=M+1}^{M+N} \sum_{\nu=M+1}^{M+N} ?[r]c*_i^\mu? \braket { \chi_{\mu} | \chi_{\nu} } ?[l]d^{\nu}_j?
	  \end{multlined}\\
	 &= 0 
\end{split}
\end{equation*}
The second and third terms are identically zero since $\braket { \chi_{\mu} | \chi_{\nu} } = 0$.
In the limit of orbital localisation such that a spin orbital describes either molecule A or B only,  $?[r]c*_i^\mu?  = 0$ for the first term and $?[l]d^{\nu}_j? = 0$ or the fourth term (Table \ref{table: Vanishing coefficients}), removing each of those terms.

\subsubsection{Overlap and Hamiltonian matrices}
We derive in Appendix A.1 and A.2 expressions for the Overlap and Hamiltonian matrix elements between states of the two molecules A and B. The Overlap matrix can be written as:
\begin{equation}
	\braket { U | V } =  \braket { U^{A} | V^{A} } \braket { U^{B} | V^{B} } 
\end{equation}
From the expression, we see that the overlap matrix is \emph{multiplicatively separable}.\\ 
In matrix form, this corresponds to:
\begin{equation}
	\boldsymbol{S}_{AB} =  \boldsymbol{S}_{A} \otimes \boldsymbol{S}_{B}
\end{equation}
which reflects the use of a direct product basis.\\
The Hamiltonian matrix can be expressed as:
\begin{equation}
\begin{aligned}
\begin{split}
	\braket { U | \hat{H} | V } ={ }& \braket {U^{A} | \hat{H} | V^{A} }\braket { U^{B} | V^{B} } \\
	 & + \braket {U^{B} | \hat{H} | V^{B} }\braket { U^{A} | V^{A} }  
\end{split}
\end{aligned}
\end{equation}
In contrast to the Overlap matrix, the Hamiltonian matrix is \emph{additively separable}.\\ 
In matrix form, this corresponds to 
\begin{equation}
	\boldsymbol{H}_{AB} = \boldsymbol{H}_{A} \otimes \boldsymbol{S}_{B} + \boldsymbol{S}_{A} \otimes \boldsymbol{H}_{B}
\end{equation}
In the special case (orthogonal basis states) where $\boldsymbol{S}_{A} = \boldsymbol{1}_{A}$ and $\boldsymbol{S}_{B} = \boldsymbol{1}_{B}$, the equation reduces to the more familiar form\cite{Helgaker}:
\begin{equation}
	\boldsymbol{H}_{AB} = \boldsymbol{H}_{A} \otimes \boldsymbol{1}_{B} + \boldsymbol{1}_{A} \otimes \boldsymbol{H}_{B}
\end{equation}

\subsubsection{Size-Extensivity}
The generalised eigenvalue equations for the two molecules A and B individually are:
\begin{equation}
	\boldsymbol{H}_{A} \boldsymbol{D}_{A}  = \boldsymbol{S}_{A} \boldsymbol{D}_{A} E_{A}
\end{equation}
\begin{equation}
	\boldsymbol{H}_{B} \boldsymbol{D}_{B}  = \boldsymbol{S}_{B} \boldsymbol{D}_{B} E_{B},
\end{equation}
and these define NOCI wavefunctions, and energies for the separated molecular systems.
We now take $\boldsymbol{D}_{AB} = \boldsymbol{D}_{A} \otimes \boldsymbol{D}_{B}$ as a trial wavefunction of the combined systems using the direct product basis. In the combined system, the generalised eigenvalue equation is:
\begin{equation}
	\boldsymbol{H}_{AB} \boldsymbol{D}_{AB} = \boldsymbol{S}_{AB} \boldsymbol{D}_{AB} E_{AB}
\end{equation}
Expanding the LHS and comparing the expression, we get:
\begin{equation}
\begin{aligned}
\begin{split}
	\boldsymbol{H}_{AB} \boldsymbol{D}_{AB} = { }& (\boldsymbol{H}_{A} \otimes \boldsymbol{S}_{B} + \boldsymbol{S}_{A} \otimes \boldsymbol{H}_{B}) (\boldsymbol{D}_{A} \otimes \boldsymbol{D}_{B}) \\
	= { }& (\boldsymbol{H}_{A} \otimes \boldsymbol{S}_{B})(\boldsymbol{D}_{A} \otimes \boldsymbol{D}_{B}) \\
	& + (\boldsymbol{S}_{A} \otimes \boldsymbol{H}_{B})(\boldsymbol{D}_{A} \otimes \boldsymbol{D}_{B}) \\
	= { }& (\boldsymbol{S}_{A} \boldsymbol{D}_{A} \otimes \boldsymbol{S}_{B} \boldsymbol{D}_{B})(E_{A} + E_{B}) \\
	= { }&\boldsymbol{S}_{AB} \boldsymbol{D}_{AB} (E_{A} + E_{B})
\end{split} 
\end{aligned}
\end{equation}
Hence, we can conclude that $E_{AB} = E_{A} + E_{B}$. By localising spin orbitals on each molecule in a dimer, we have demonstrated formally that the NOCI approach will yield size-extensive results. 

\section{Results and Discussion}
	\subsection{Computational Details}
All NOCI and CCSD calculations were performed using Q-Chem 5.3\cite{QChem} and sa-CASSCF calculations were performed using OpenMolcas \cite{OpenMolcas}. In-house Python codes were used to perform spin rotation on the sb-UHF states and to determine the irreducible representations spanned by the NOCI states found. Quantum Toolbox in Python (QuTiP) 4.3.1\cite{QuTiP} was employed to simultaneously diagonalise representation matrices. 3D-rendering of molecules were produced with the IQMol software. \\
The general approach for performing NOCI are as follows:
\begin{enumerate}
\item  Run SCF metadynamics\cite{Thom1, Thom2} on a given molecule [Q-Chem]
\item  Extract the lowest energy UHF solution (sb-UHF) and generate spin-rotated states from it [Python]
\item Read in the generated states and run NOCI [Q-Chem]
\item Perform symmetry analysis on the obtained NOCI states [Python]
\end{enumerate}

\subsection{Collinear Hydrogen Dimer} \label{h2dimer}
\begin{Figure}
\centering
\includegraphics[scale=0.4]{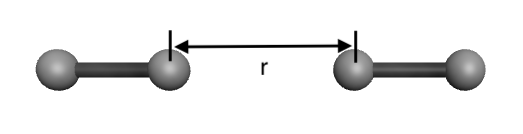}
\captionof{figure}{Collinear Hydrogen, $r=10\si{\angstrom}$}
\label{fig:Collinear Hydrogen}
\end{Figure}

We will consider the case where $r=10\si{\angstrom}$ (Figure \ref{fig:Collinear Hydrogen}). 

\subsubsection{Spin-flip case}
\begin{Figure}
\centering
\includegraphics[scale=0.33]{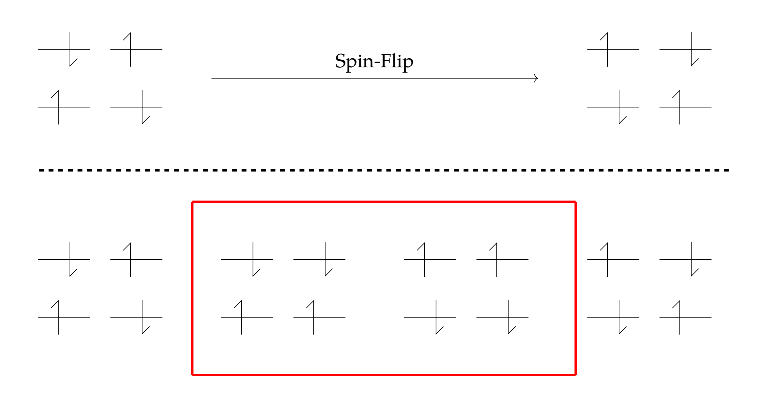}
\captionof{figure}{(Top) The two configurations obtained by applying spin-flip of a sb-UHF solution describing the dimer. (Bottom) The four configurations required to obtain a $M_{s} = 0$ Singlet and Triplet state for the dimer. The two configurations enclosed in the box will not be obtainable from the spin-flip as shown in the top figure.}
\label{fig:H2_configurations}
\end{Figure}

The method of spin-rotation can be more intuitively understood in the limiting case where $\theta = \pi$. When $\theta = \pi$ the spin-rotation operator is equivalent to the time-reversal operator, $\hat{\mathcal{T}}$. The effect of the time-reversal operator is as follows:
\begin{equation}
\begin{split}
\hat{\mathcal{T}} \ket{\alpha} = \ket{\beta} \\
\hat{\mathcal{T}} \ket{\beta} = -\ket{\alpha}
\end{split}
\end{equation}
It therefore behaves as a spin-flip operator, albeit with an additional phase factor. Operating on a sb-UHF solution as described in Figure \ref{fig:H2_configurations} (Top Left), we find two possible configurations related by spin-flip.\\
However, four different spin configurations (Figure \ref{fig:H2_configurations}, bottom) are required for the dimer (Each hydrogen molecule can have two possible sb-UHF states). Therefore, performing a global spin-flip on the whole hydrogen dimer will not yield sufficient configurations to give a size-extensive description of the hydrogen dimer. \\
This is shown in Table \ref{table: Size-extensivity}, where the global spin rotation approach gives an energy 45$mE_{h}$ above the size consistent local spin rotation approach.

\pagebreak
%\resizebox{\columnwidth}{!}{
\begin{tabular}{ c c c } 
 \hline \hline
 & Spin-Rotations & NOCI Energy/ $E_{h}$  \\ 
 \hline
Molecule    				&	-		&   -0.9934465141 \\
\\
Dimer 			&  No rotation 		&   -1.9414513430  \\
 &  $\hat{R}(\pi \hat{ \textbf{y}})$&\\
\\
Dimer   	&  No rotation		&	 -1.9868905053  \\
& $\hat{R}_{A}(\pi \hat{ \textbf{y}})$ &\\
& $\hat{R}_{B}(\pi \hat{ \textbf{y}})$ &\\
& $\hat{R}(\pi \hat{ \textbf{y}})$ &\\
\hline \hline
\end{tabular} 
%}
\captionof{table}{Ground state NOCI energies of the hydrogen molecule and hydrogen dimer. $\hat{R}_{A}(\pi \hat{ \textbf{y}})$ and $\hat{R}_{B}(\pi \hat{ \textbf{y}})$ refers to localised spin-rotations a single hydrogen molecule in the dimer, denoted $A$ and $B$ respectively. $\hat{R}(\pi \hat{ \textbf{y}})$ refers to spin-rotation on the entire dimer (both hydrogen molecules). These spin-rotations are taken to be of angle $\pi$ about the y axis. Inclusion of localised spin-rotation is found to give size-consistent energies.}
\label{table: Size-extensivity}

\subsubsection{General spin-rotation about an axis}
We can extend the ideas in the previous subsection by using more spin rotations. The set of rotations $\{ \hat{R} (\frac{n\pi}{2} \hat{ \textbf{y}} ) | n = 1, 2, 3, 4\}$ on each hydrogen molecule produces nine NOCI states from sixteen spin-rotated states. Denoting the coupled state from two separate molecules $A$ and $B$ with spin quantum numbers $m$ and $n$ respectively as $A^{m}B^{n}$, we expect there to be an upper limit of 16 states formed from $A^{0}B^{0}$, $A^{1}B^{0}$, $A^{0}B^{1}$ and $A^{1}B^{1}$ using Clebsch-Gordan coupling (Table \ref{table: Clebsch-Gordan}). \\
Given that we have found 3 states for a single molecule (see Supporting Information), we would expect $3^{2}$ possible states in the corresponding dimer. These 9 states were indeed found (Table \ref{table: y-axis-H2_2}). \\

{
\centering
\begin{tabular}{ c c } 
 \hline \hline
 Spin states & Possible spins (S)  \\ 
 \hline
$A^{0}B^{0}$    &   0 \\
$A^{0}B^{1}$,  $A^{1}B^{0}$ &   1  \\
$A^{0}B^{2}$,   $A^{2}B^{0}$   &   2  \\
$A^{1}B^{1}$    &   0, 1, 2  \\
$A^{1}B^{2}$,  $A^{2}B^{1}$    &   1, 2, 3 \\
$A^{2}B^{2}$    &   0, 1, 2, 3, 4  \\
\hline \hline
\end{tabular} 
\captionof{table}{Possible spins (S) for each spin state. This is found by using the Clebsch-Gordan series. For a spin state $A^{m}B^{n}$, $S$ takes on the values $|m+n|, |m+n-1|, ..., |m-n|$}
\label{table: Clebsch-Gordan}
} 
\vspace{\baselineskip}
There are three unique energies found across the 9 states.
State 1 is unique, States 2-5 are degenerate, and so are States 6-9. Comparing the values of NOCI energy in Table \ref{table: y-axis-H2_2} and that of molecular hydrogen (Supporting Information), we find that the NOCI energy is size-extensive, as we expect. \\

%\begin{table}
%\centering
\begin{tabular}{ c c c c c } 
 \hline \hline
 States & $S^{2}$ &  $\langle S_{y}^{2} \rangle$ & $\langle S_{y} \rangle$ &NOCI Energy\\ %& Eigenket representation\\
 \hline
1   &  0.000&        0.000& 0.000&-1.98689058\\ %& $\ket{0,0}\ket{0,0}$ \\
\hline
2  &  2.000&        1.000& 0.000&-1.88403040\\ %& $\frac{1}{\sqrt{2}}(\ket{0,0}\ket{1,1} + \ket{0,0}\ket{1,-1})$ \\
\hline
3  &  2.000&       1.000& 0.000&-1.88403040\\ %& $\frac{1}{\sqrt{2}}(\ket{0,0}\ket{1,1} - \ket{0,0}\ket{1,-1})$\\
\hline
4  &  2.000&        1.000& 0.000&-1.88403040\\ %& $\frac{1}{\sqrt{2}}(\ket{1,1}\ket{0,0} + \ket{1,-1}\ket{0,0})$\\
\hline
5  &  2.000&        1.000&  0.000&-1.88403040\\ %& $\frac{1}{\sqrt{2}}(\ket{1,1}\ket{0,0} - \ket{1,-1}\ket{0,0})$\\
\hline
6  &  2.000&        0.000& 0.000&-1.78116950\\ %& $\frac{1}{\sqrt{2}}(\ket{1,1}\ket{1,-1} + \ket{1,-1}\ket{1,1})$\\
\hline
7   &  6.000&        4.000&  0.000&-1.78116950\\ %& $\frac{1}{\sqrt{2}}(\ket{1,1}\ket{1,1} + \ket{1,-1}\ket{1,-1})$\\
\hline
8  &  6.000&        4.000& 0.000& -1.78116950\\ %& $\frac{1}{\sqrt{2}}(\ket{1,1}\ket{1,1} - \ket{1,-1}\ket{1,-1})$\\
\hline
9  &  2.000&         0.000& 0.000&-1.78116950\\ %5& $\frac{1}{\sqrt{2}}(\ket{1,1}\ket{1,-1} - \ket{1,-1}\ket{1,1})$\\
\hline \hline
\end{tabular}
\captionof{table}{SR-NOCI states of collinear $H_{2}$ dimer found by spin rotation about the y-axis. A maximum of 9 ($3^{2}$) states can be found as there were only 3 states found from SR-NOCI on the hydrogen molecule.}
\label{table: y-axis-H2_2}
%\end{table}
\begin{table*}
\centering
\begin{tabular}{ c c c c c c } 
 \hline \hline
 States & $S^{2}$ &  $\langle S_{z}^{2} \rangle$ & $\langle S_{z} \rangle$ & NOCI Energy & Eigenket representation\\
\hline
1  &   0.000  &   0.000   &    0.000   &   -1.98689058 & $\ket{0,0}\ket{0,0}$\\
\hline
2  &   2.000  &   1.000  &   1.000   &    -1.88403040 & $\frac{1}{\sqrt{2}}  (\ket{0,0}\ket{1,1} + \ket{1,1}\ket{0,0})$\\
\hline
3   &   2.000  &   1.000  &   -1.000   &    -1.88403040 & $\frac{1}{\sqrt{2}}  (\ket{0,0}\ket{1,-1} + \ket{1,-1}\ket{0,0})$\\  
\hline
4   &   2.000   &  1.000  &   -1.000   &    -1.88403040 & $\frac{1}{\sqrt{2}}  (\ket{0,0}\ket{1,-1} - \ket{1,-1}\ket{0,0})$\\
\hline
5   &   2.000  &   1.000  &   1.000   &    -1.88403040 & $\frac{1}{\sqrt{2}}  (\ket{0,0}\ket{1,1} - \ket{1,1}\ket{0,0})$\\
\hline
6   &   2.000  &   0.000  &   0.000   &    -1.88403040 & $\frac{1}{\sqrt{2}}  (\ket{0,0}\ket{1,0} + \ket{1,0}\ket{0,0})$\\
\hline
7   &   2.000  &   0.000  &   0.000   &    -1.88403040 & $\frac{1}{\sqrt{2}}  (\ket{0,0}\ket{1,0} - \ket{1,0}\ket{0,0})$\\
\hline
8   &   2.000  &  1.000  &   -1.000   &    -1.78116950 & $\frac{1}{\sqrt{2}}  (\ket{1,0}\ket{1,-1} - \ket{1,-1}\ket{1,0})$\\
\hline
9   &   2.000 &   1.000  &   1.000   &    -1.78116950 & $\frac{1}{\sqrt{2}}  (\ket{1,1}\ket{1,0} - \ket{1,0}\ket{1,1})$\\
\hline
10    &   6.000  &   1.000  &   -1.000   &    -1.78116950 & $\frac{1}{\sqrt{2}} (\ket{1,0}\ket{1,-1} + \ket{1,-1}\ket{1,0})$\\
\hline
11    &   6.000  &   1.000 &   1.000   &    -1.78116950 & $\frac{1}{\sqrt{2}} (\ket{1,1}\ket{1,0} + \ket{1,0}\ket{1,1})$\\
\hline
12    &   6.000  &   4.000  &    2.000   &   -1.78116950 & $\ket{1,1}\ket{1,1}$\\
\hline
13    &   2.000  &   0.000  &    0.000   &   -1.78116950 & $\frac{1}{\sqrt{2}}  (\ket{1,1}\ket{1,-1} - \ket{1,-1}\ket{1,1})$\\
\hline
14    &   6.000  &   4.000  &    -2.000   &   -1.78116950 & $\ket{1,-1}\ket{1,-1}$\\
\hline
15   &    0.000  &   0.000  &   0.000   &    -1.78116950 & $\frac{1}{\sqrt{3}} (\ket{1,1}\ket{1,-1} + \ket{1,-1}\ket{1,1} - \ket{1,0}\ket{1,0})$\\
\hline
16    &   6.000  &   0.000 &   0.000   &    -1.78116950 & $\frac{1}{\sqrt{6}} (\ket{1,1}\ket{1,-1} + \ket{1,-1}\ket{1,1} + 2\ket{1,0}\ket{1,0})$\\
\hline \hline
\end{tabular}
\captionof{table}{NOCI states of collinear $H_{2}$ dimer found by spin rotation about the x- and y-axes. All 16 possible spin states are found. The degeneracy of the solutions are 1,6 and 9, corresponding to the spin states $A^{0}B^{0}$, $A(B)^{0}B(A)^{1}$ and $A^{1}B^{1}$.}
\label{table: xy-axis-H2_2}
\end{table*}

\subsection{Spin rotation about two axes}
Although we were unable to obtain all of the possible spin states by rotation about one axis, the correct spin states can be found with two rotation axes.
Performing rotations with this set of rotations: \\
$\{ \hat{R} (\frac{n\pi}{2} \hat{ \textbf{y}} ) | n = 1, 2, 3, 4  \} \cup \{ \hat{R} (\frac{n\pi}{2} \hat{ \textbf{x}} ) | n = 1, 2, 3, 4  \}$, we find all 16 states (Table \ref{table: xy-axis-H2_2}). 
The NOCI energies found have degeneracies in the ratio of 1:6:9, as we expect.  We also see that the full possible 16 states have been found with the correct $S^{2}$ and $S_{z}$ values. This demonstrates that the method is able to project out the various spin symmetry respecting components of the symmetry broken solution, and points to a possible way forward.\\

\begin{figure}[H]
\centering
  \begin{tikzpicture}[scale=0.55, every node/.style={scale=0.55}]
\begin{axis}[
	width=16cm,
	height=9cm,
    xlabel={\textbf{C-C distance/ $\si{\angstrom}$}},
    ylabel={\textbf{SCF Energy/ $E_{h}$}},
    xmin=1.20, xmax=3.4,
    ymin=-77.1, ymax=-76.6,
    xtick={1.2, 1.5,2,2.5,3, 3.4},
    ytick={-77.1, -77.0, -76.9, -76.8, -76.7, -76.6},
    every axis plot/.append style={ultra thick},
    legend style={at={(0.50,-0.2)},anchor=north,legend columns=5, nodes={scale=1.2, transform shape}, /tikz/every even column/.append style={column sep=0.5cm}},
]
  
\addplot[
	%UHF
    color=Green,
    mark=None,
    ]
    table [x index=0, y index=1]{Ethene_Minimal_Basis_auxiliary_data/UHF.dat};
    
\addplot[
	%hUHF
	only marks,
    color=Green,
    mark=o,
    ]
    table [x index=0, y index=1]{Ethene_Minimal_Basis_auxiliary_data/hUHF.dat};
    
\addplot[
	%RHF
    color=Orange,
    mark=None,
    dashed,
    ]
    table [x index=0, y index=1]{Ethene_Minimal_Basis_auxiliary_data/RHF.dat};
    
\addplot[
	%ALMO
    color=Red,
    mark=None,
    dotted,
    ]
    table [x index=0, y index=4]{Ethene_Minimal_Basis_auxiliary_data/ALMO.dat};
    
\addplot[
	%ALMO 0
    color=MidnightBlue,
    mark=None,
    loosely dotted,
    ]
    table [x index=0, y index=1]{Ethene_Minimal_Basis_auxiliary_data/ALMO_0.dat};

   \addlegendentry{UHF}
   \addlegendentry{hUHF}
   \addlegendentry{RHF}
   \addlegendentry{ALMO ($M_{s} = \pm 1$)}
   \addlegendentry{ALMO ($M_{s} = 0$)}
\end{axis}
\end{tikzpicture}
\caption{Plot of the UHF and (Unrestricted) ALMO solutions across various C-C bond distances using the STO-3G basis. These solutions are spin contaminated. A Coulson-Fischer point exists under 1.30$\si{\angstrom}$, where the UHF solution turns holomorphic. The ALMO solutions are continuous across all C-C bond distances, however.}
\label{fig: UHF Plots}
\end{figure}
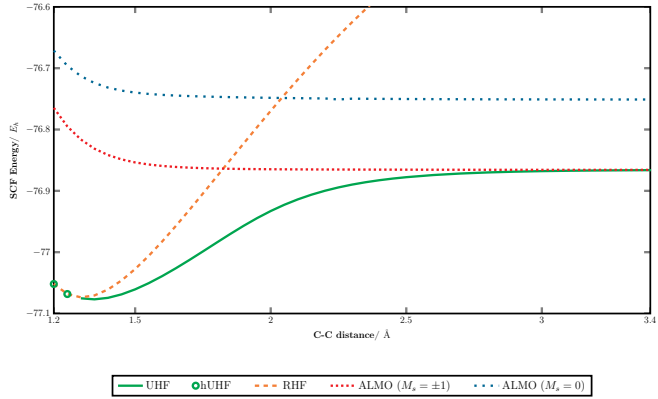

\subsection{Number of spin-rotations}
It might be tempting to use more spin-rotations in the hopes of achieving better results. However, we note that the using more spin-rotated states (larger basis) will generally lead to overcompleteness of basis when some of these states are linearly dependent. While equation (\ref{NOCI}) takes care of such linear dependencies automatically, it will increase computational cost. For an $N$ electron system the maximum spin will be $S_{max}=N/2$, so using $S_{max}^{2}$ linearly independent spin-rotated states on a monomer (and therefore $S_{max}^{4}$ on a dimer) would generate all the spin-pure NOCI states. In practice the very high $S$ states are very high in energy and do not significantly contaminate the UHF states which are reference states for SR-NOCI. Therefore, we simply increase the number of monomer spin-rotations (using the square of this for the dimer) sufficiently such that the NOCI energies of the spin-states we are concerned with do not change appreciably and that their $S^2$ values indicate spin symmetry has been sufficiently restored (typically within 0.01 of an eigenvalue of $S^2$).

%For a given molecule, there are $L^{2}$ spin states (therefore $L^{4}$ for a dimer) where $L$ is the maximum orbital angular momentum number. Therefore, $L^{2}$ linearly independent spin-rotated states should generate all the spin pure NOCI states, in principle. In practice, however, we do not use $L^{2}$ spin-rotated states out of computational cost considerations. We simply increase the number of spin-rotated states sufficiently such that the NOCI energies of the spin-states we are concerned with do not change appreciably and that their spin symmetry has been restored. 

\section{Stretched ethene}
Using spin rotated states as a basis, we studied the dissociation of an ethene molecule into two carbenes. We modelled the dissociation process by elongating the C-C bond whilst leaving remaining bond lengths and bond angles constant. This was studied using both STO-3G and 6-31G* bases to investigate the effect of bases and to understand the correlation captured in our method.

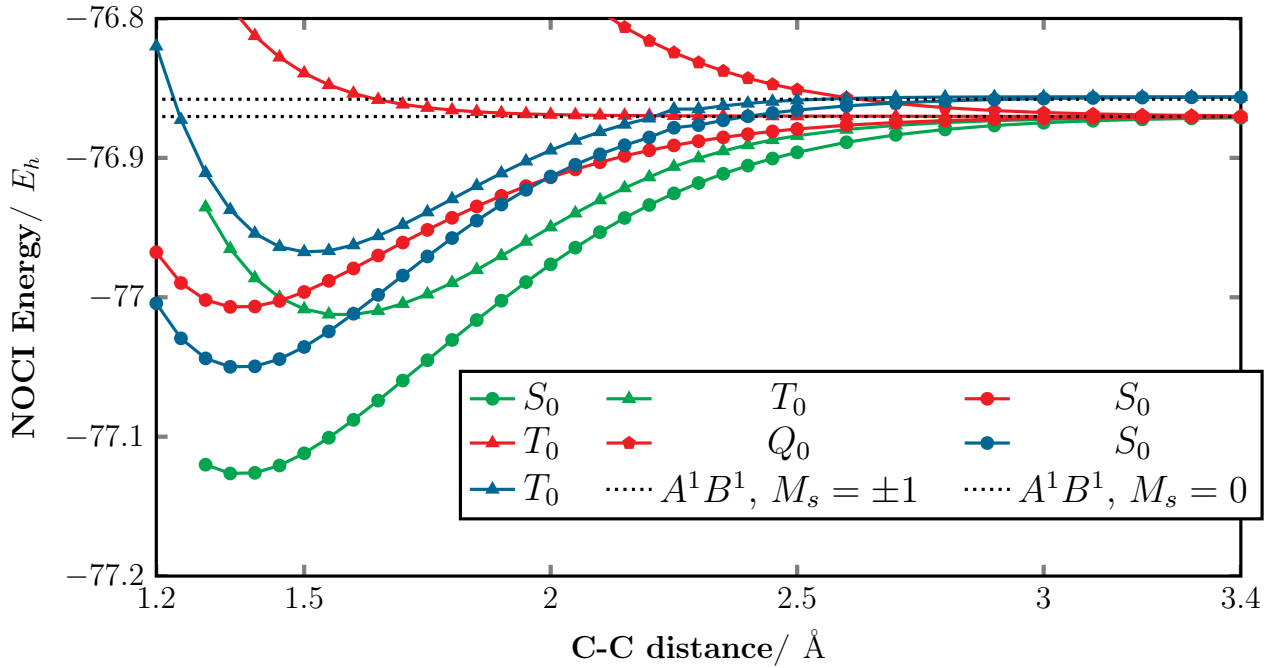
\begin{figure*}
  \centering
\begin{tikzpicture}
\begin{axis}[
	width=16cm,
	height=9cm,
    xlabel={\textbf{C-C distance/ $\si{\angstrom}$}},
    ylabel={\textbf{NOCI Energy/ $E_{h}$}},
    xmin=1.2, xmax=3.4,
    ymin=-77.2, ymax=-76.8,
    xtick={1.2,1.5,2,2.5,3,3.4},
    ytick={-77.2,-77.1, -77.0, -76.9, -76.8},
    legend style={at={(0.65,0.37)},anchor=north,legend columns=3, nodes={scale=1.2, transform shape}, /tikz/every even column/.append style={column sep=0.5cm}},
]
    
%UHF

\addplot[
	%S0
    color=Green,
    mark=*,
    ]
    table [x index=0, y index=1]{Ethene_Minimal_Basis_NOCI_data/datafile/singletdata.dat};
    
\addplot[
	%T0
    color=Green,
    mark=triangle*,
    ]
    table [x index=0, y index=1]{Ethene_Minimal_Basis_NOCI_data/datafile/tripletdata.dat};

%M_{s} = 1/-1 ALMO
  
\addplot[
	%S0
    color=Red,
    mark=*,
    ]
    table [x index=0, y index=1]{Ethene_Minimal_Basis_ALMO_data/datafile/singletdata.dat};
    
\addplot[
	%T0
    color=Red,
    mark=triangle*,
    ]
    table [x index=0, y index=1]{Ethene_Minimal_Basis_ALMO_data/datafile/tripletdata.dat};
    
\addplot[
	%Q0
    color=Red,
    mark=pentagon*,
    ]
    table [x index=0, y index=1]{Ethene_Minimal_Basis_ALMO_data/datafile/quintetdata.dat};
    
%M_{s} = 0 ALMO

\addplot[
	%S0
    color=MidnightBlue,
    mark=*,
    ]
    table [x index=0, y index=1]{Ethene_Minimal_Basis_ALMO_0_data/datafile/singletdata.dat};
    
\addplot[
	%T0
    color=MidnightBlue,
    mark=triangle*,
    ]
    table [x index=0, y index=1]{Ethene_Minimal_Basis_ALMO_0_data/datafile/tripletdata.dat};
    
	\addplot[mark=none, black, dotted] {-76.87032503};
	\addplot[mark=none, black, dotted] {-76.85792699};
	
   \addlegendentry{$S_{0}$}
   \addlegendentry{$T_{0}$}
   \addlegendentry{$S_{0}$}
   \addlegendentry{$T_{0}$}
   \addlegendentry{$Q_{0}$}
   \addlegendentry{$S_{0}$}
   \addlegendentry{$T_{0}$}
   \addlegendentry{$A^{1}B^{1}$, $M_{s} = \pm 1$}
   \addlegendentry{$A^{1}B^{1}$, $M_{s} = 0$}

\end{axis}
\end{tikzpicture}
\caption{SR-NOCI states from the UHF solution (green), a pair of $M_{s} = \pm 1$ ALMO solutions (red) and a pair of $M_{s} = 0$ ALMO solutions (blue).  $X = S_{0},T_{0}$ and $Q_{0}$ refers to the ground state Singlet, Triplet and Quintet states respectively. The dotted lines represents the asymptotic energies of the spin states. At larger C-C bond distances, the energy of the spin states tend towards the asymptotic energies, demonstrating that SR-NOCI is a size consistent method.}
\label{fig: Ethene Dissociation}
\end{figure*}

\subsection{Canonical UHF orbitals}

\begin{figure*}
\begin{tikzpicture}
\begin{groupplot}[
    group style={
        group name=my plots,
        group size = 2 by 2,
        xlabels at=edge bottom,
        xticklabels at=edge bottom,
        vertical sep=0pt,
        horizontal sep=2cm
    },
    width=8cm,
    height=7cm,
    xlabel=\textbf{C-C distance/ $\si{\angstrom}$},
    xmin=1.2, xmax=3.4,
    xtick={1.2,1.5,2,2.5,3, 3.4},
    tickpos=left,
    ytick align=outside,
    xtick align=outside,
    every axis plot/.append style={ultra thick},
    legend style={at={(-0.25,-0.25)},anchor=north,legend columns=5, nodes={scale=1, transform shape}, /tikz/every even column/.append style={column sep=0.5cm}}
]

\nextgroupplot[ymin=-77.28, ymax=-76.85, ytick={ -77.2, -77.1, -77.0, -76.9}, ylabel ={\textbf{Energy/ $E_{h}$}} ]
\addplot[
	%SR-NOCI
    color=Orange,
    dashed,
    ]
    table [x index=0, y index=1]{Ethene_Minimal_Basis_NOCI_data/datafile/singletdata.dat};
    
%\addplot[
	%SR-NOCI
%    color=Red,
%    dashed,
%    ]
%    table [x index=0, y index=1]{Ethene_Minimal_Basis_NOCI_data/global/singletdata.dat};
    
\addplot[
	%CASSCF
    color=Fuchsia,
    dotted,
    ]
    table [x index=0, y index=1]{Ethene_Minimal_Basis_auxiliary_data/CASSCF.dat};

\addplot[
	%CCSD
    color=MidnightBlue,
    loosely dotted,
    ]
    table [x index=0, y index=1]{Ethene_Minimal_Basis_auxiliary_data/CCSD.dat};
    
 \addplot[
 	%UHF
 	color=Green,
 	]
 	table [x index=0, y index=1]{Ethene_Minimal_Basis_auxiliary_data/UHF.dat};

%%%%%%%%%%%%%%%%%

\nextgroupplot[ymin=-78.35, ymax=-77.80, ytick={-78.3, -78.1, -77.9}, ylabel ={\textbf{Energy/ $E_{h}$}} ]
\addplot[
	%SR-NOCI
    color=Orange,
    dashed,
    ]
    table [x index=0, y index=1]{Ethene_6-31Gstar_data/SR_NOCI_Localised/singletdata.dat};
    
%\addplot[
	%SR-NOCI
%    color=Red,
%    dashed,
%    ]
%    table [x index=0, y index=1]{Ethene_6-31Gstar_data/SR_NOCI_Global/singletdata.dat};
    
\addplot[
	%CASSCF
    color=Fuchsia,
    dotted,
    ]
    table [x index=0, y index=1]{Ethene_6-31Gstar_data/Ethene_CASSCF_data.dat};

\addplot[
	%CCSD
    color=MidnightBlue,
    loosely dotted,
    ]
    table [x index=0, y index=1]{Ethene_6-31Gstar_data/Ethene_CCSD_data.dat};
    
 \addplot[
 	%UHF
 	color=Green,
 	]
 	table [x index=0, y index=1]{Ethene_6-31Gstar_data/UHF.dat};

%%%%%%%

\nextgroupplot[ymin=-0.3, ymax=0.05, ytick={-0.40, -0.30, -0.20, -0.10 ,  0}, ylabel={\textbf{Relative Energy/ $E_{h}$}} ]
\addplot[
	%SR-NOCI
    color=Orange,
    dashed
    ]
    table [x index=0, y expr=\thisrowno{1}+76.87032808]{Ethene_Minimal_Basis_NOCI_data/datafile/singletdata.dat};
    
%\addplot[
	%SR-NOCI
%    color=Red,
%    dashed
%    ]
%    table [x index=0, y expr=\thisrowno{1}+76.86584551]{Ethene_Minimal_Basis_NOCI_data/global/singletdata.dat};
    
\addplot[
	%CASSCF
    color=Fuchsia,
    dotted,
    ]
    table [x index=0,  y expr=\thisrowno{1}+76.94206534]{Ethene_Minimal_Basis_auxiliary_data/CASSCF.dat};

\addplot[
	%CCSD
    color=MidnightBlue,
    loosely dotted,
    ]
    table [x index=0, y expr=\thisrowno{1}+76.94171566]{Ethene_Minimal_Basis_auxiliary_data/CCSD.dat};
    
 \addplot[
 	%UHF
 	color=Green,
 	]
 	table [x index=0, y expr=\thisrowno{1}+76.86584551]
{Ethene_Minimal_Basis_auxiliary_data/UHF.dat};

		\addplot[mark=none, black, dotted] {0};
   
 %%%%%%%%%%%

\nextgroupplot[ymin=-0.3, ymax=0.05, ytick={-0.30, -0.20, -0.10 ,  0}, ylabel={\textbf{Relative Energy/ $E_{h}$}} ]\addplot[
	%SR-NOCI
    color=Orange,
    dashed,
    ]
    table [x index=0, y expr=\thisrowno{1}+77.83637303]{Ethene_6-31Gstar_data/SR_NOCI_Localised/singletdata.dat};
    
%\addplot[
	%SR-NOCI
%    color=Red,
%    dashed,
%    ]
%    table [x index=0, y expr=\thisrowno{1}+77.83210068]{Ethene_6-31Gstar_data/SR_NOCI_Global/singletdata.dat};
    
\addplot[
	%CASSCF
    color=Fuchsia,
    dotted,
    ]
    table [x index=0,  y expr=\thisrowno{1}+77.90182366]{Ethene_6-31Gstar_data/Ethene_CASSCF_data.dat};

\addplot[
	%CCSD
    color=MidnightBlue,
    loosely dotted
    ]
    table [x index=0, y expr=\thisrowno{1}+78.0312417]{Ethene_6-31Gstar_data/Ethene_CCSD_data.dat};
    
 \addplot[
 	%UHF
 	color=Green,
 	]
 	table [x index=0, y expr=\thisrowno{1}+77.83210068]
{Ethene_6-31Gstar_data/UHF.dat};

	\addplot[mark=none, black, dotted] {0};
   
   \addlegendentry{SR-NOCI (UHF)}
   \addlegendentry{sa-CASSCF}
   \addlegendentry{CCSD}
   \addlegendentry{UHF}
   
\end{groupplot}

\end{tikzpicture}
\caption{(Top) Absolute energies for the SR-NOCI, sa-CASSCF(12, 12), CCSD and UHF solutions. sa-CASSCF averages the two lowest energy solution found, while CCSD takes the R=UHF solution tracked from C-C distance of $1.2 \si{\angstrom}$. The UHF solution is added for comparison. (Bottom) Dissociation curve is plotted relative to twice the energy of a triplet methylene molecule found from the respective methods. In particular, CASSCF(6, 6) was performed on the methylene molecule. (Left) Using STO-3G basis. (Right) Using 6-31G* basis.}
\label{fig: Benchmark}
\end{figure*}
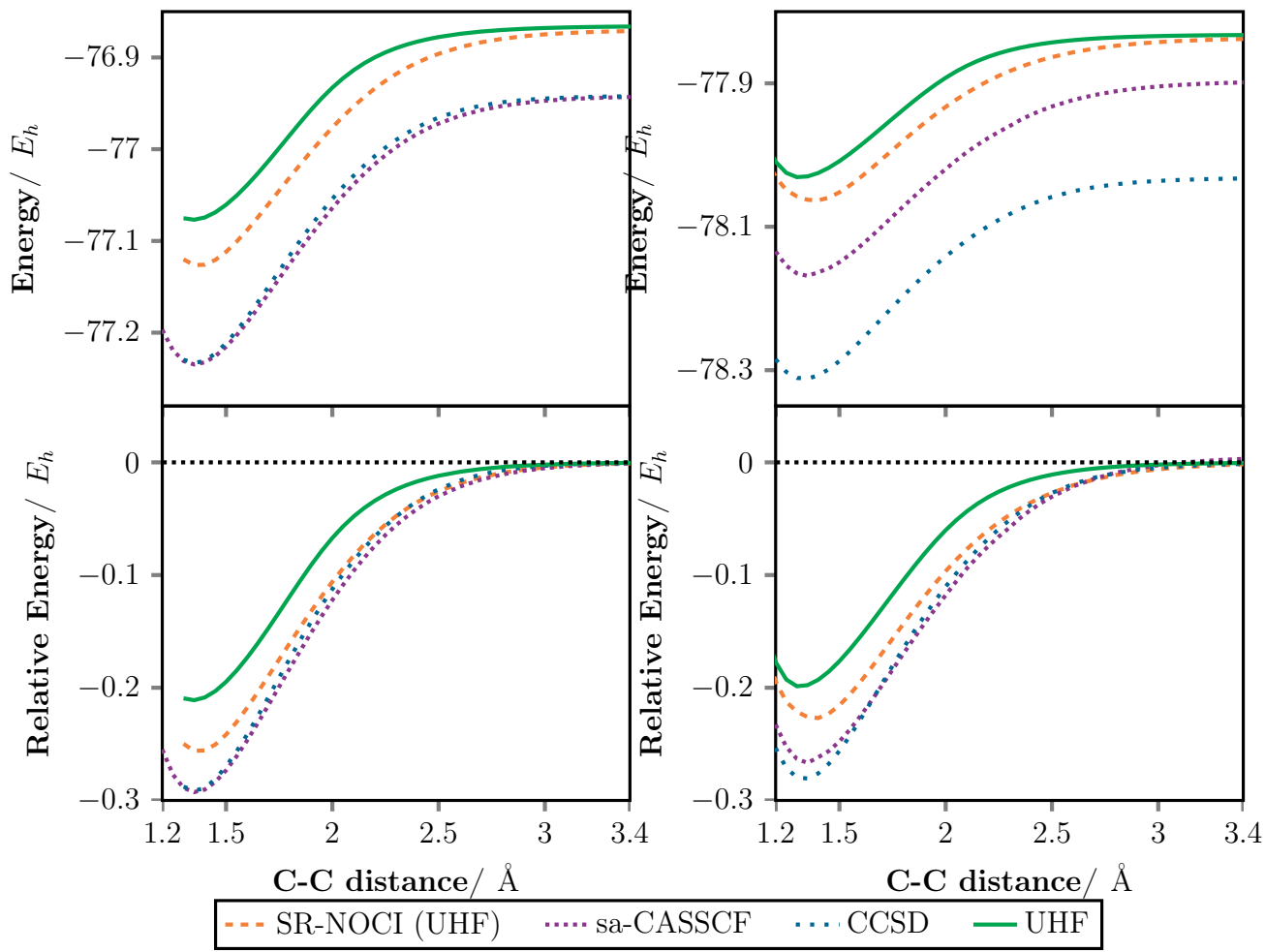

With the lowest energy sb-UHF solution of ethene found by metadynamics (Figure \ref{fig: UHF Plots}), SR-NOCI was applied to the solutions at each C-C bond distance. Through SR-NOCI, the low energy Singlet and Triplet states are recovered (Figure \ref{fig: Ethene Dissociation}) at all C-C bond distances. At dissociation, the NOCI solution is shown to be size consistent. Using the STO-3G basis, both the Singlet and Triplet states' energies approach the limiting value of $-76.87033 E_{h}$, corresponding to the dissociated product, a pair of $M_{s} = \pm 1$ carbenes. This demonstrates the utility of SR-NOCI in modelling dissociation processes while maintaining the important properties of spin symmetry and size consistency.\\
The absolute energy found using the SR-NOCI method is shown to be an overestimate of the energies when compared to methods such as state-averaged CASSCF (sa-CASSCF) and CCSD (Figure \ref{fig: Benchmark}). We attribute the difference in absolute energies to the lack of dynamic correlation captured by our treatment. This claim is substantiated by observing the effects of changing the basis from STO-3G to 6-31G*. The addition of polarisation functions in the basis sets allows for the inclusion of more dynamic correlation effects, as can be observed by the increased absolute energy gap between sa-CASSCF and CCSD curves (Figure \ref{fig: Benchmark}, Top Right), while the energy difference between SR-NOCI and sa-CASSCF curves remains relatively constant.\\
We have also plotted the relative energy curve for the various methods, where the relative energy for each method is defined as the difference between the absolute energy and that of two independent triplet methylene molecules at infinite separation. All methods used (sa-CASSCF, CCSD and SR-NOCI) display size consistency in this example (Figure \ref{fig: Benchmark}, Bottom). The size inconsistency in each method is tabulated (Table \ref{table: Size-inconsistency}), which highlights the relative size consistency of the SR-NOCI method.\\
A drawback with the use of UHF solutions is that they may disappear at Coulson-Fischer points. Without the corresponding UHF states as reference, discontinuities in the NOCI energies will arise. As such, the disappearance of UHF solutions along the dissociation process is concerning for our proposed method. It is possible to analytically continue the UHF solution into the complex plane via holomorphic SCF\cite{HoloHF1, HoloHF2, HoloHF2}. For example, the ground state UHF solution for ethene coalesces with the RHF solution below the C-C bond distance of $1.3 \si{\angstrom}$, but holomorphic SCF can be used to find the corresponding holomorphic-UHF solutions (Figure \ref{fig: UHF Plots}). While these holomorphic states can in principle be used for SR-NOCI, we do not have a code for it at present. Instead, we shall demonstrate that Absolutely Localised Molecular Orbitals (ALMO) can also be used to produce an alternate set of NOCI solutions that are also well-behaved at all C-C bond distances.\\
\newline
\begin{tabular}{ c c c } 
 \hline \hline
							& STO-3G/$mE_h$ 			& 6-31G*/$mE_h$  \\ 
 \hline
SR-NOCI    			&	0.007						&   0.02 \\
\\
sa-CASSCF 			&  0.014 						&   5.37  \\
\\
CCSD   					&  0.008						&	 0.03  \\
\hline \hline
\end{tabular} 
\captionof{table}{The size inconsistency of the three methods: SR-NOCI, sa-CASSCF and CCSD for STO-3G and 6-31G* basis. The size inconsistency is defined as the absolute difference in energy of ethene molecule at a C-C bond distance of $10 \si{\angstrom}$ and that of two $M_s = 1$ methylene molecules calculated by the same method. All values given in milli-Hartrees.}
\label{table: Size-inconsistency}

\subsection{Absolutely Localised Molecular Orbitals (ALMO)}
Absolutely Localised Molecular Orbitals (ALMOs)\cite{ALMO1, ALMO2} can be utilised to address the problem of disappearing UHF solutions. Using the ALMO SCF method, the orbitals in a UHF state are always localised on one of the two molecules, even when they are very close.  For this system, we were able to track the lowest energy solution across all C-C bond distances and therefore provides a method not hindered by coalescing UHF solutions. We note that the ALMOs centred on each molecule are not orthogonal to each other and at larger separations between molecules, the ALMO solutions become equivalent to the conventional localised UHF solutions.\\
We first trace the lowest energy sb-UHF solutions across C-C distances between $1.2 \si{\angstrom}$ to $3.5 \si{\angstrom}$, and the results are plotted in Figure \ref{fig: UHF Plots}. Treating the ALMO solutions found at each geometry using SR-NOCI, we have found a series of spin states, and the low-lying Singlet, Triplet and Quintet states have been plotted in Figure \ref{fig: Ethene Dissociation}. The plot confirms the size consistency of the NOCI states, as the NOCI energies approaches the sum of the NOCI energies on the individual carbene fragments. The size consistent property extends to excited states as well.\\
Figure \ref{fig: Ethene Dissociation} puts the equilibrium C-C bond length at around $1.35 \si{\angstrom}$. We note that the experimental C-C bond length for ethene is $1.33 \si{\angstrom}$, which is in good agreement with the calculated value. 
We note that this is likely to be fortuitous as SR-NOCI does not account for dynamic correlation and that localised orbitals may not be a good basis to capture the bonding between carbene fragments.\\
While the use of ALMO has allowed us to side-step the problem of coalescing solutions, the localised nature of the wavefunction necessitates that the ground state energy found would be higher than that of the UHF one by the variational principle. As such, while SR-NOCI can introduce binding to the ALMO solutions, the binding energy remains an underestimate compare to that found using sb-UHF solution as the reference. Another common problem with ALMO would be the difficulty in describing charge-transfer state which are important for heterolytic bond breaking, but that is not a problem in our present study. Therefore, we encourage the use of sb-UHF solutions as reference whenever possible.\\

\section{Comparison between MS-NOCI and SR-NOCI}
In previous work done by the group, various SCF solutions were used as a basis for NOCI. We will compare the number of solutions used, energy and $\langle S^{2} \rangle$ of the NOCI ground state found to demonstrate that the spin-rotated basis is a good alternative.\\
We shall use fluorine, $D_{4h}$ cyclobutadiene, and Alizarin donor and acceptor states which were previously used as model systems in research carried out by the group. A common set of rotations: $\{ \hat{R} (\frac{n\pi}{4} \hat{ \textbf{y}} ) | n = 1, 2, ..., 8  \} \cup \{ \hat{R} (\frac{n\pi}{4} \hat{ \textbf{x}} ) | n = 1, 2, ..., 8  \}$ is used across the systems for consistency, unless stated otherwise. The basis set used is consistent with those used in previous works. Table 8 summarises the difference in results between SR-NOCI against the previous method of MS-NOCI. 

\subsection{Fluorine}
Fluorine is well-known as a challenging system for computational chemistry due to the high degree of both static and dynamic correlation. $E_{min}$ of Fluorine found with the spin-rotated approach is in good agreement with MS-NOCI using 3 SCF states (Differing by $7.8 \times 10^{-4} E_{h}$). This is suggests that the spin-rotated approach simply recovers static correlation.\\ %Table number is hardcoded here
For $F_{2}$ at $2 \si{\angstrom}$, we observe that SR-NOCI gives a higher energy ground state, suggesting that the use of one symmetry-broken UHF state is insufficient in capturing the dynamic correlation in Fluorine. As the bond length of $F_{2}$ increases to $8 \si{\angstrom}$, however, we see that the present method gives a slightly lower energy. This agrees with our understanding as the dynamic correlation between electrons on each F atom is made less significant at this bond length.\\

\subsection{Cyclobutadiene}
\begin{figure*}
  \centering

  \begin{tikzpicture}
\begin{axis}[
	width=16cm,
	height=9cm,
    xlabel={\textbf{Distortion angle/ Degrees}},
    ylabel={\textbf{NOCI Energy/ $E_{h}$}},
    xmin=-17, xmax=17,
    ymin=-153.75, ymax=-153.5,
    xtick={-15, -10, -5, 0, 5, 10, 15 },
    ytick={-153.75, -153.70, -153.65, -153.60, -153.55, -153.50},
    every axis plot/.append style={ultra thick},
    legend style={at={(0.5,-0.2)},anchor=north,legend columns=5, nodes={scale=1.2, transform shape}, /tikz/every even column/.append style={column sep=0.5cm}},
    legend entries={$^{1}B_{1g}$, $^{3}A_{2g}$, $^{1}A_{1g}$, $^{1}B_{2g}$, SA-CASSCF,
    						  $^{1}B_{1g}$, $^{3}A_{2g}$, $^{1}A_{1g}$, $^{1}B_{2g}$, MS-NOCI,
    						  $^{1}B_{1g}$, $^{3}A_{2g}$, $^{1}A_{1g}$, $^{1}B_{2g}$, SR-NOCI,}
]

%%% Reference data %%%

\addplot[
	%1B1g
	color=Turquoise,
    mark=*,
    ]
    table [x index=0, y index=1]{Reference_cyclobutadiene/sa-cas_4-4_singlet.dat};
    
\addplot[
	%3A2g
    color=Turquoise,
    mark=triangle*,
    ]
    table [x index=0, y index=1]{Reference_cyclobutadiene/sa-cas_4-4_triplet.dat};
    
\addplot[
	%1B1g
    color=Turquoise,
    mark=square*,
    ]
    table [x index=0, y index=2]{Reference_cyclobutadiene/sa-cas_4-4_singlet.dat};
    
\addplot[
	%1B1g
    color=Turquoise,
    mark=pentagon*,
    ]
    table [x index=0, y index=3]{Reference_cyclobutadiene/sa-cas_4-4_singlet.dat};
    
\addlegendimage{color=Turquoise}
    
 %%% MS-NOCI data %%%
 
 \addplot[
	%1B1g
    color=Green,
    mark=*,
    ]
    table [x index=0, y index=1]{Reference_cyclobutadiene/noci_E.dat};
    
\addplot[
	%3A2g
    color=Green,
    mark=triangle*,
    ]
    table [x index=0, y index=2]{Reference_cyclobutadiene/noci_E.dat};
    
\addplot[
	%1A1g
    color=Green,
    mark=square*,
    ]
    table [x index=0, y index=3]{Reference_cyclobutadiene/noci_E.dat};
    
\addplot[
	%1B2g
    color=Green,
    mark=pentagon*,
    ]
    table [x index=0, y index=4]{Reference_cyclobutadiene/noci_E.dat};

\addlegendimage{color=Green}
    
%%% SR-NOCI %%%

\addplot[
	%1B1g
    color=Fuchsia,
    mark=*,
    ]
    table [x index=0, y index=1]{Cyclobutadiene1_data/datafile/singletdata.dat};
    
\addplot[
	%3A2g
    color=Fuchsia,
    mark=triangle*,
    ]
    table [x index=0, y index=1]{Cyclobutadiene1_data/datafile/tripletdata.dat};

\addplot[
	%1A1g
    color=Fuchsia,
    mark=square*,
    ]
    table [x index=0, y index=1]{Cyclobutadiene_auxiliary_data/singletdata.dat};

\addplot[
	%1B2g
    color=Fuchsia,
    mark=pentagon*,
    ]
    table [x index=0, y index=1]{Cyclobutadiene3_data/datafile/singletdata.dat};
    
\addlegendimage{color=Fuchsia}

\end{axis}
\end{tikzpicture}
\caption{Purple: SR-NOCI data,  Green: MS-NOCI data and Turquoise: SA-CASSCF data. Energies of ground and excited states of cyclobutadiene at various geometries. $^{1}B_{1g}$ and $^{3}A_{2g}$ states were found using UHF1, $^{1}A_{1g}$ state using both UHF2 solutions (degenerate at distortion angle of zero) and $^{1}B_{2g}$ with UHF3.}
\label{fig:Cyclobutadiene}
\end{figure*}
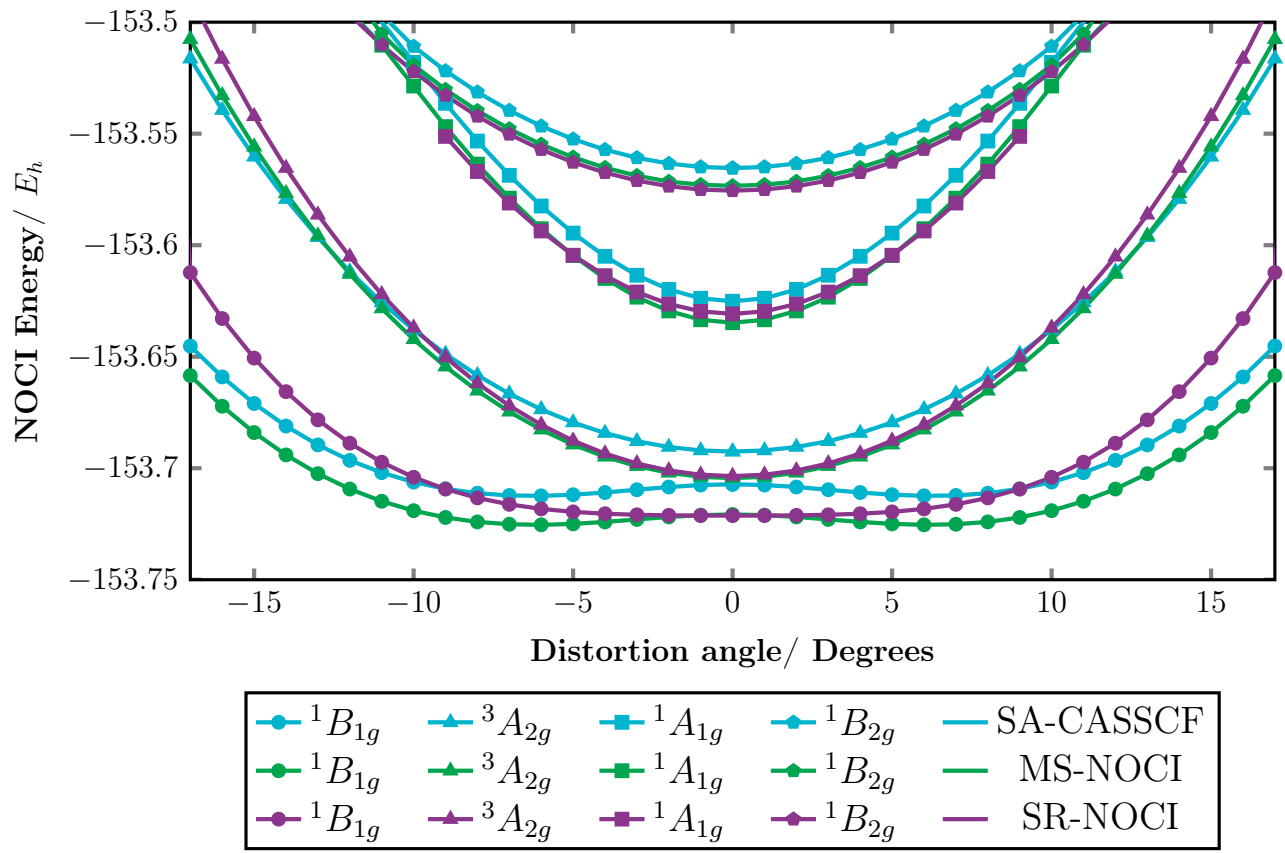

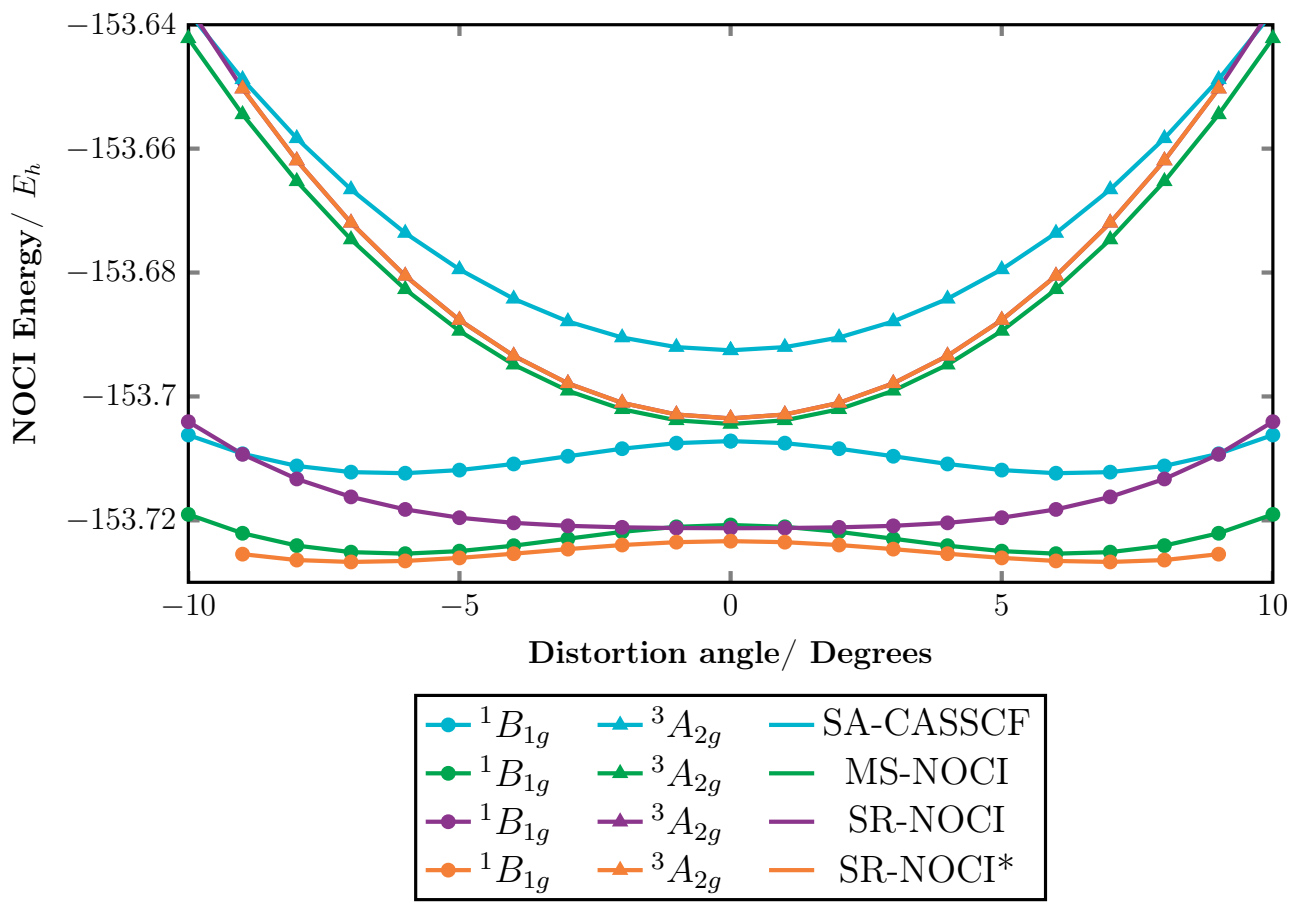
\begin{figure*}
  \centering

  \begin{tikzpicture}
\begin{axis}[
	width=16cm,
	height=9cm,
    xlabel={\textbf{Distortion angle/ Degrees}},
    ylabel={\textbf{NOCI Energy/ $E_{h}$}},
    xmin=-10, xmax=10,
    ymin=-153.73, ymax=-153.64,
    xtick={-10, -5, 0, 5, 10},
    ytick={-153.72, -153.70, -153.68, -153.66, -153.64},
    every axis plot/.append style={ultra thick},
    legend style={at={(0.5,-0.2)},anchor=north,legend columns=3, nodes={scale=1.2, transform shape}, /tikz/every even column/.append style={column sep=0.5cm}},
    legend entries={$^{1}B_{1g}$, $^{3}A_{2g}$, SA-CASSCF,
    						  $^{1}B_{1g}$, $^{3}A_{2g}$, MS-NOCI,
    						  $^{1}B_{1g}$, $^{3}A_{2g}$, SR-NOCI,
    						  $^{1}B_{1g}$, $^{3}A_{2g}$, SR-NOCI*}
]

%%% Reference data %%%

\addplot[
	%1B1g
	color=Turquoise,
    mark=*,
    ]
    table [x index=0, y index=1]{Reference_cyclobutadiene/sa-cas_4-4_singlet.dat};
    
\addplot[
	%3A2g
    color=Turquoise,
    mark=triangle*,
    ]
    table [x index=0, y index=1]{Reference_cyclobutadiene/sa-cas_4-4_triplet.dat};
    
\addlegendimage{color=Turquoise}
    
 %%% MS-NOCI data %%%
 
 \addplot[
	%1B1g
    color=Green,
    mark=*,
    ]
    table [x index=0, y index=1]{Reference_cyclobutadiene/noci_E.dat};
    
\addplot[
	%3A2g
    color=Green,
    mark=triangle*,
    ]
    table [x index=0, y index=2]{Reference_cyclobutadiene/noci_E.dat};

\addlegendimage{color=Green}

	%%% SR-NOCI data %%%
	
\addplot[
	%1B1g
    color=Fuchsia,
    mark=*,
    ]
    table [x index=0, y index=1]{Cyclobutadiene1_data/datafile/singletdata.dat};
    
\addplot[
	%3A2g
    color=Fuchsia,
    mark=triangle*,
    ]
    table [x index=0, y index=1]{Cyclobutadiene1_data/datafile/tripletdata.dat};
    
\addlegendimage{color=Fuchsia}

\addplot[
	%1B1g
    color=Orange,
    mark=*,
    ]
    table [x index=0, y index=1]{Cyclobutadiene_auxiliary_data/singlet_extensive.dat};
    
\addplot[
	%3A2g
    color=Orange,
    mark=triangle*,
    ]
    table [x index=0, y index=1]{Cyclobutadiene_auxiliary_data/triplet_extensive.dat};
    
 \addlegendimage{color=Orange}

\end{axis}
\end{tikzpicture}
\caption{Expansion of Figure \ref{fig:Cyclobutadiene} with added SR-NOCI solutions. SR-NOCI* (orange) refers to the NOCI states found using UHF1 and both UHF2 solutions in a single SR-NOCI calculation. The other solutions are the same as in Figure \ref{fig:Cyclobutadiene}. The $^{1}B_{1g}$ curve found in this manner has the double-well shape as found in both SA-CASSCF and MS-NOCI methods.  $^{3}A_{2g}$ state remains unchanged, however.}
\label{fig:SR-NOCI-Cyclobutadiene}
\end{figure*}

For $D_{4h}$ Cyclobutadiene, several sb-UHF states can be found, and we shall denote the lowest energy sb-UHF state as UHF1, second lowest as UHF2, and so on.  The two lowest energy SR-NOCI states ($^{1}B_{1g}$ and $^{3}A_{2g}$) were found from applying spin rotation to the UHF1, while the $^{1}B_{2g}$ SR-NOCI state was found by applying spin-rotation to UHF3. UHF2, however, comprises a pair of spatial symmetry broken states (Figure \ref{fig:sb-UHF}). They are degenerate at a distortion angle of zero where cyclobutadiene has $D_{4h}$ symmetry and becomes non-degenerate upon distortion of the molecule. Using only one of these symmetry broken solutions as reference will fail to provide an accurate representation of the SR-NOCI $^{1}A_{1g}$ state. As such, both the degenerate UHF2 states were tracked across the molecular geometries and used as reference for SR-NOCI in a single SR-NOCI calculation.\\
The SR-NOCI results are overlayed with the results previously reported by our group\cite{Hugh-NOCI} (Figure \ref {fig:Cyclobutadiene}). We find that the NOCI energies found from SR-NOCI are comparable to that found through MS-NOCI and sa-CASSCF approaches. This lends justification to our claim that SR-NOCI can provide a balanced description of both ground and excited states through the use of different UHF states as the reference state for SR-NOCI.\\
It should be remarked that the $^{1}A_{1g}$ state is abruptly terminated at distortion angle of $\pm 9$ degrees. This is due to one of the UHF2 solutions turning holomorphic past this angle (Figure \ref{fig:sb-UHF}). As the use of holomorphic states is not the objective of this paper, we have omitted the tracking of the $^{1}A_{1g}$ state beyond this distortion angle and will be left for future work.\\
We also observe that there is a notable difference in shape (parabolic) of $^{1}B_{1g}$ solution curve produced by SR-NOCI as compared to SA-CASSCF and MS-NOCI (double-well). This can be attributed to the use of only UHF1, which itself has a parabolic solution curve, as reference for SR-NOCI. We therefore see that in cases with significant multi-reference character, the use of a single solution for SR-NOCI cannot capture the full complexity of the solution. To investigate this further, UHF1 and both the UHF2 solutions were used as references in a single SR-NOCI calculation. The use of the two UHF2 solutions was motivated by the fact that each of their solution curves is a parabola with minima on each side of the zero distortion angle. The addition of these two UHF2 solutions should thus enable the description of a double-welled ground state. This was found to be the case (Figure \ref{fig:SR-NOCI-Cyclobutadiene}). It can be seen that upon the inclusion of UHF2 solutions, the $^{1}B_{1g}$ solution curve gives two minima, and the curve is comparable to that found in MS-NOCI. No change, however, was observed for the $^{3}A_{2g}$ state as the triplet states found with UHF2 are too high in energy to interact appreciably with those found from UHF1 ($~0.1 E_{h}$). Hence, there is no change in triplet NOCI energy going from that found with UHF1 to that found with the combination of UHF1 and UHF2 solutions.\\
This example illustrates the possible deficiencies in using a single sb-UHF solution for SR-NOCI in all cases. As a general guideline, we propose that one should first use metadynamics to determine the presence of degenerate solutions. If they exist, one should instead apply SR-NOCI to the degenerate set of solutions, rather than any single sb-UHF solution.\\

\begin{table*}
\begin{threeparttable}[t]
\centering
\scalebox{0.8}{
\begin{tabular}{ *{7}{c} }
    \hline \hline
Molecule    & \multicolumn{3}{c}{MS-NOCI}
            & \multicolumn{3}{c }{SR-NOCI} \\
    \hline
       &   Dimension of NOCI\tnote{a}   &   NOCI Energy  &  $\langle S^{2} \rangle$ &  Dimension of NOCI\tnote{b} &   NOCI Energy  &  $\langle S^{2} \rangle$\\
   \hline
$F_{2}$\tnote{*} ($E_{min}$)\tnote{$\dagger$}	&	3	&	-198.73848	& 0.002 &	14	&	-198.73770  & 0.000 \\      
    \hline
$F_{2}$\tnote{$\ddagger$}  ($2 \si{\angstrom}$)   &   8  &  -198.77008   &  0.007  &   14  &   -198.76155 &  0.000\\
    \hline
$F_{2}$\tnote{$\ddagger$}  ($8 \si{\angstrom}$)   &   8    &  -198.75051 &  0.007   &   14    &   -198.75497   & 0.000 \\
    \hline
$D_{4h}$ Cyclobutadiene\tnote{$\ddagger$}   &  12     &   -153.72075 ($^{1}B_{1g}$)   &  0.080 &  14    &    -153.72127  ($^{1}B_{1g}$)   &  0.000\\
										  &      12     &       { }{ }{ }{ }{ }0.01630\tnote{c}   ($^{3}A_{2g}$)      &   2.006    &  14 &          { }{ }{ }{ }{ }0.01773\tnote{c}   ($^{3}A_{2g}$)   &  2.000\\
										  &      12     &        { }{ }{ }{ }{ }0.06987\tnote{c}   ($^{1}A_{1g}$)      &   0.039    &   28\tnote{d}  &          { }{ }{ }{ }{ }0.09046\tnote{c}   ($^{1}A_{1g}$)  &  0.000\\
										  &      12     &        { }{ }{ }{ }{ }0.12331\tnote{c}   ($^{1}B_{2g}$)       &   0.000    &   14 &         { }{ }{ }{ }{ }0.14570\tnote{c}    ($^{1}B_{2g}$)   &  0.000\\
\hline
Alizarin-Titanium (D)\tnote{$\mathsection$}   &   30    &   -1984.69793\tnote{e}    &  0.659   &  14    &    -1984.70549    & 0.004 \\
    \hline
Alizarin-Titanium (A)\tnote{$\mathsection$}   &   30    &   -1984.69457\tnote{e}    &   0.810  &  14    &    -1984.71074   & 0.067 \\
    \hline \hline
\end{tabular}
}
\caption{Comparison of data found using MS-NOCI against SR-NOCI}

\begin{tablenotes}
     \item[a] The dimensions is found from the number of HF solutions used. The previous method uses\\
     multiple HF solutions for NOCI. Whenever the solution used is a UHF one (UHF states with \\
     $M_{s} = 0$ are assumed), its spin-flipped partner is used as well.
     \item[b] The dimensions is found from the number of spin rotations used. The dimension of 14 arises\\
     from the following set of rotations: $\{ \hat{R} (\frac{n\pi}{4} \hat{ \textbf{y}} ) | n = 1, 2, ..., 8  \} \cup \{ \hat{R} (\frac{n\pi}{4} \hat{ \textbf{x}} ) | n = 1, 2, ..., 8  \}$. \\
     The reason for having 14 states when there are 16 rotations is because $\hat{R} (\pi \hat{ \textbf{y}}) = \hat{R} (\pi \hat{ \textbf{x}})$ and \\
     $\hat{R} (2\pi \hat{ \textbf{y}}) = \hat{R} (2\pi \hat{ \textbf{x}}) = \hat{I}$, $\hat{I}$ being the identity operator.
     \item[c] These energies corresponds to the excitation energy from their respective ground $^{1}B_{1g}$ states.
     \item[d] The dimension is doubled as both degenerate solutions were used in spin-rotation to restore\\
     spatial symmetry.
     \item[e] The energies here are slightly different from those previously reported. The discrepancy was\\
     due to the lack of phase correction after L\"{o}wdin Pairing in the previous Q-Chem code.\\
     This has since been rectified.
     \item[*] 6-31G basis set was used.
     \item[$\dagger$] $E_{min}$ is the lowest (MS or SR) NOCI energy found across the F-F bond geometries sampled. 
     \item[$\ddagger$] cc-pVDZ basis was used.
     \item[$\mathsection$] 6-31G* basis set was used.
   \end{tablenotes}
\end{threeparttable}
\label{tab:Comparison}
\end{table*}

\begin{Figure}
\centering
\includegraphics[scale=0.25]{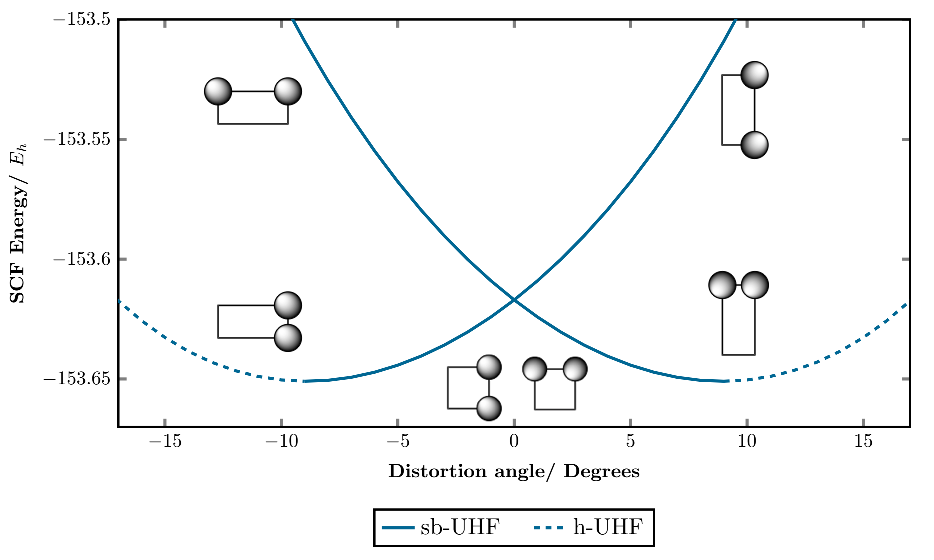}
\captionof{figure}{Illustration of the spatial symmetry breaking in the second lowest energy UHF solution (denoted UHF2 in main text) of cyclobutadiene at various distortion angles. The two solutions are degenerate when distortion angle is zero ($D_{4h}$ symmetry).}
\label{fig:sb-UHF}
\end{Figure}

\subsection{Alizarin-Titanium Complex}
The value of the spin-rotated approach can be more fully appreciated when we move on to larger systems. In larger systems, it is common to find many sb-UHF solutions and hence the selection of chemically relevant states is a non-trivial task. As with the example of both forms of the Alizarin-Titanium complex (Donor and Acceptor states, as defined in a previous work \cite{Kris}), the NOCI states found with MS-NOCI usually suffer from spin contamination. Using SR-NOCI, we have found singlet ground states in each of the Alizarin-Titanium complexes which are lower in energy than previously reported. The improved spin purity of the NOCI states is heartening as it gives us more confidence that we are describing the correct spin states.\\
Previously, 30 SCF states were used, as compared to the current approach uses only 14 states to obtain a lower energy ground state which respects spin symmetry. This can therefore also be useful in calculating the spin states of large molecular systems which may be too computationally intensive with currently available methods.\\

\section{Conclusion}
We have conducted a study on the viability of using spin-rotated states as a basis for NOCI (SR-NOCI). Using this method, the energies found were comparable to those found with SCF metadynamics states as a basis (MS-NOCI). Our current approach comes with three additional advantages: 
\begin{enumerate}
\item It is a simple and generic method where the same methodology can be used for molecular systems of various sizes, albeit with a change in number of spin rotations required.
\item In the case of non-degeneracy, only one sb-UHF state is required to obtain SR-NOCI states with static correlation recovered and possess good spin quantum numbers.
\item It offers the advantage of having a smaller dimension of NOCI in larger systems.
\end{enumerate}
This method, however, only recovers the static correlation as it projects out the various spin components of a sb-UHF state.  A further improvement can be made by employing the NOCI-PT2 approach recently published\cite{NOCIPT2} by our group to account for the dynamic correlation.\\
We have also applied our method to the dissociation of an ethene molecule into two carbenes using both canonical UHF orbitals and ALMO. The results show that the dissociation is size consistent and the states respect spin symmetry at each step of the dissociation. \\
By applying spin-rotation to individual molecules of the system and performing NOCI on the resulting states, this approach sets itself apart from the Projected Hartree Fock approach, which to the best of our knowledge have not been applied on individual fragments. Through the use of fragments, we also achieve size-consistency and good spin quantum numbers. This is also different from the State Averaged Resonating Hartee Fock (SA ResHF) approach proposed by Nite and Jimenez-Hoyos\cite{Nite} as we do not optimise the determinants used as the basis for NOCI and no state-averaging is required.\\
We can see a potential application of this work in Singlet Fission modelling\cite{Smith1, Smith2, Casanova, Havenith1, Havenith2, Zimmerman}. An important state in Singlet Fission is the correlated triplet state, $^{1}(TT)$, as denoted in literature\cite{1TT}. This corresponds to a $A^{1}B^{1}$ system of $S=0$ in our notation. We are also able to track the spin state as the molecular components $A$ and $B$ separate. This will enable us to follow the dissociation of the correlated triplet pair state into two independent triplets. The balanced description of both ground and excited state in our approach is beneficial to obtaining more accurate excitation energies, which is another important aspect of Singlet Fission research. We hope to apply this method to the modelling of these photochemical processes in a future work.

\section{Acknowledgements}
We would like to thank Dr. Hugh Burton and Dr. Bang Huynh for their helpful advice and discussions in the preparation of this manuscript.

\section{Notes}
The Supporting Information is available free of charge on the ACS Publications website.
Research data supporting this work and further information can be found at on the University of Cambridge's online repository.

\appendix
\section{Appendix}

\subsection{Rotation matrix}
We use spin-rotation about an axis to illustrate the similarity between our approach and using Wigner D-matrices. From the rotation matrix we defined in equation (\ref{RotationMatrix}),
\begin{equation*}
\begin{split}
	\boldsymbol{R} &= \begin{pmatrix}
								 		\cos\frac{\theta}{2} - in_{z}\sin\frac{\theta}{2} & (-n_{y} - in_{x})\sin\frac{\theta}{2} \\
								 		(n_{y} - in_{x})\sin\frac{\theta}{2}  & \cos\frac{\theta}{2} + in_{z}\sin\frac{\theta}{2}
							   \end{pmatrix} \\
\end{split}
\end{equation*}
we set $n_{x} = n_{z} = 0$ and $n_{y} = 1$ (Rotate about the y-axis). This reduces to
\begin{equation}
\begin{split}
	\boldsymbol{R} &= \begin{pmatrix}
								 		\cos\frac{\theta}{2} & -\sin\frac{\theta}{2} \\
								 		\sin\frac{\theta}{2}  & \cos\frac{\theta}{2}
							   \end{pmatrix} \\
\end{split}
\label{SimplifiedRotationMatrix}
\end{equation}
We can compare this to the use of a projection operator as defined in another work\cite{PHF}. $\beta$ refers to an angle in the set of Euler angles $(\alpha, \beta, \gamma)$, following the Z-Y-Z convention. $d_{mm}^{s}(\beta)$ is the Wigner little D-matrix.
\begin{equation}
\hat{P}_{mm}^{s} = \frac{2s+1}{2} \int_{0}^{\pi} d\beta \sin\beta\: d_{mm}^{s}(\beta) \: \mathrm{e}^{i\beta \hat{S}_{y}}
\end{equation}
We discretise this expression, turning the integral into a summation:
\begin{equation}
\begin{split}
\hat{P}_{mm}^{s} &= \frac{2s+1}{2} \sum_{\beta} \sin\beta\: d_{mm}^{s}(\beta) \: \mathrm{e}^{i\beta \hat{S}_{y}}\\
						   &= \frac{2s+1}{2} \sum_{\beta} \sin\beta\: d_{mm}^{s}(\beta) \: \begin{pmatrix}
								 																									\cos\frac{\beta}{2}  & \sin\frac{\beta}{2} \\
								 																									-\sin\frac{\beta}{2}  & \cos\frac{\beta}{2}
							   																									  \end{pmatrix} \\
\end{split} 
\end{equation}
Taking only two points such that $\beta = 0$ and $\beta = -\theta$, we obtain:
\begin{equation}
\hat{P}_{mm}^{s} = -\frac{2s+1}{2}  \sin\frac{\theta}{2} \: d_{mm}^{s}\Big(\frac{\theta}{2}\Big) \: \begin{pmatrix}
								 																									\cos\frac{\theta}{2} & -\sin\frac{\theta}{2} \\
								 																									\sin\frac{\theta}{2}  & \cos\frac{\theta}{2}
							   																									  \end{pmatrix} \\
\label{WignerExpression}
\end{equation}
It is clear that equations (\ref{SimplifiedRotationMatrix}) and (\ref{WignerExpression}) differ only by a constant factor. In practical applications the projection operator is discretised and therefore these two approaches are equivalent. To further support our claim, we have performed a calculation on the $H_{2}$ dimer (Section \ref{h2dimer}) using the Wigner D-matrices using 11 grid points evenly spaced between angles $\beta = 0$ and $\beta = \pi$. The numerical integration was performed with the Simpson's method via Python's Scipy implementation. The results of the calculation was compared against a SR-NOCI calculation with \emph{global} spin-rotations of $\{ \hat{R} (\frac{2n\pi}{5} \hat{ \textbf{y}} ) | n = 0, 1, 2, 3, 4\}$ (more spin-rotations do not improve the result any further as the basis of spin-rotated states become overcomplete) in table 8. From the table, we observe that both the methods give equivalent results for the energies of both $S_{0}$ and $T_{0}$ states. The description of $S_{1}$ state by SR-NOCI is poor but can be easily improved by introducing spin-rotations about a different axis (e.g. x-axis) to expand the basis of linearly independent spin-rotated states.\\

\begin{tabular}{ c c c } 
 \hline \hline
 & Wigner-D matrix/ $E_{h}$  &  SR-NOCI (Global rotation)/ $E_{h}$  \\ 
 \hline
$S_{0}$				&	-1.96135029		&   -1.96135029 \\
$T_{0}$		   	    &  -1.88403036 		&   -1.88403036 \\
$S_{1}$				&  -1.78438300		&	 -1.78116950  \\
\hline \hline
\label{Comparison}
\end{tabular}
\captionof{table}{Comparison of the energies of spin-states of the collinear hydrogen dimer found using Wigner D-matrix method and the SR-NOCI method. $S_{0}$ and $S_{1}$ refers to the ground and first-excited singlet states, respectively, while $T_{0}$ refers to the ground triplet state. This example demonstrates numerically the equivalence of both methods when we are using only global spin-rotations with SR-NOCI.}

\begin{figure*}[h!]
\centering
\begin{tikzpicture}[
style1/.style={
  matrix of math nodes,
  every node/.append style={text width=#1,align=center,minimum height=5ex},
  nodes in empty cells,
  left delimiter=[,
  right delimiter=],
  },
style2/.style={
  matrix of math nodes,
  every node/.append style={text width=#1,align=center,minimum height=5ex},
  nodes in empty cells,
  left delimiter=[,
  right delimiter=],
  }
]
\matrix[style1=0.45cm] (1mat)
{
  & & & & & \\
  & & & & & \\
  & & & & & \\
  & & & & & \\
  & & & & & \\
  & & & & & \\
};

\draw[]
  (1mat-3-1.south west) -- (1mat-3-6.south east);
\draw[]
  (1mat-1-3.north east) -- (1mat-6-3.south east);
\node[font=\huge] 
  at ([xshift=-10pt]1mat-2-2) {$U^{A}$};
\node[font=\huge] 
  at (1mat-2-5) {$0$};
\node[font=\huge] 
  at (1mat-5-5.east) {$U^{B}$};
\node[font=\huge] 
  at (1mat-5-2) {$0$};
  
\draw[decoration={brace,mirror,raise=12pt},decorate]
  (1mat-1-1.north west) -- 
  node[left=15pt] {$M$} 
  (1mat-3-1.south west);
\draw[decoration={brace,mirror,raise=12pt},decorate]
  (1mat-4-1.north west) -- 
  node[left=15pt] {$N$} 
  (1mat-6-1.south west);

\draw[decoration={brace,raise=7pt},decorate]
  (1mat-1-1.north west) -- 
  node[above=8pt] {$P$} 
  (1mat-1-3.north east);
\draw[decoration={brace,raise=7pt},decorate]
  (1mat-1-4.north west) -- 
  node[above=8pt] {$Q$} 
  (1mat-1-6.north east);

\matrix[style2=0.45cm,right=80pt of 1mat] (2mat)
{
  & & & & & \\
  & & & & & \\
  & & & & & \\
  & & & & & \\
  & & & & & \\
  & & & & & \\
};
\draw[]
  (2mat-3-1.south west) -- (2mat-3-6.south east);
\draw[]
  (2mat-1-3.north east) -- (2mat-6-3.south east);
\draw[dashed]
  (2mat-1-3.north) -- (2mat-6-3.south);
\draw[dashed]
  (2mat-1-3.north west) -- (2mat-6-3.south west);
\draw[dashed]
  (2mat-3-1.west) -- (2mat-3-6.east);
 \draw[dashed]
  (2mat-3-1.north west) -- (2mat-3-6.north east);
  
\node[font=\huge] 
  at ([xshift=-10pt]2mat-2-2) {$S^{A}$};
\node[font=\huge] 
  at (2mat-2-5) {$0$};
\node[font=\huge] 
  at (2mat-5-5.east) {$S^{B}$};
\node[font=\huge] 
  at (2mat-5-2) {$0$};
  
\draw[decoration={brace,mirror,raise=12pt},decorate]
  (2mat-1-1.north west) -- 
  node[left=15pt] {$P$} 
  (2mat-3-1.south west);
\draw[decoration={brace,mirror,raise=12pt},decorate]
  (2mat-4-1.north west) -- 
  node[left=15pt] {$Q$} 
  (2mat-6-1.south west);

\draw[decoration={brace,raise=7pt},decorate]
  (2mat-1-1.north west) -- 
  node[above=8pt] {$P$} 
  (2mat-1-3.north east);
\draw[decoration={brace,raise=7pt},decorate]
  (2mat-1-4.north west) -- 
  node[above=8pt] {$Q$} 
  (2mat-1-6.north east);
 
\draw[decoration={brace,mirror,raise=5pt},decorate]
  (2mat-6-3.south west) -- 
  node[below=7pt] {$j$} 
  (2mat-6-3.south);
  
\draw[decoration={brace,mirror,raise=12pt},decorate]
  (2mat-3-6.east) -- 
  node[right=15pt] {$i$} 
  (2mat-3-6.north east);

\end{tikzpicture}
\caption{(Left) For a state $U$ of a coupled system AB, the spin orbitals are localised on each molecule (A and B respectively) and arranged in a block diagonal fashion as shown. $U^{A(B)}$ corresponds to the spin orbitals of state $U$ centred on molecule $A(B)$. Each row represents an atomic orbital while each column represents a spin orbital. (Right) $S^{A(B)}$ is the overlap matrix centred on molecule $A(B)$. The determinant of this block matrix gives the overlap between states $U$ and $V$. The cofactor $D(i | j)$ is the determinant after a row $i$ and column $j$ has been removed. If the row and column are centred on molecule $A(B)$, $S^{B(A)}$ remains unchanged. $i$ and $j$ must correspond to atomic orbitals on the same molecule or the associated terms will be vanishing.}
\label{fig: Matrices}
\end{figure*}
\subsection{Overlap matrix}
 We emphasise that the expressions for the Overlap and Hamiltonian matrix elements presented in this and the next section rests upon the fact that molecules A and B are at infinite separation (or that their orbitals are localised). The structure of the matrices are as illustrated in Figure \ref{fig: Matrices}.
Following the exposition of L\"{o}wdin\cite{Lowdin1, Lowdin2, Lowdin3} and Mayer\cite{Mayer}, 
\begin{equation}
	U = \hat{\mathcal{A}} [u_{1}(1) ... u_{P}(P) u_{P+1}(P+1)...u_{P+Q}(P+Q)  ]
\end{equation}
\begin{equation}
	V = \hat{\mathcal{A}} [v_{1}(1) ... v_{P}(P) v_{P+1}(P+1)...v_{P+Q}(P+Q)  ]
\end{equation}
where $\hat{\mathcal{A}}$ is the anti-symmetrising operator.  The labels $P$ and $Q$ used in the Appendix are integers labelling electrons, and are not related to the arbitrary states $P$ and $Q$ in Equation 1 of the main text. After some algebra, we get the following expression for the value of the overlap between states $U$ and $V$:
\begin{equation}
\begin{aligned}
	\braket{ U | V } = & \sum_{R \in S_{P+Q}} (-1)^{r} \braket { u_{1} | v_{R_{1}} } ... \braket { u_{P} | v_{R_{P}} } \\
	& \braket { u_{P+1} | v_{R_{P+1}} } ... \braket { u_{P+Q} | v_{R_{P+Q}} }
\end{aligned}
\end{equation}
where $S_{P+Q}$ is the full symmetric group of $P+Q$ elements, and $r$ is the parity of the permutation. \\
This equation can be simplified as $\braket { u_{i} | v_{j} }  = 0$ if $u_{i}$ and $v_{j}$ are spin orbitals on different molecules. Therefore, for the set of spin orbitals $\{u_{i} | i \in 1, 2,...,P\}$, only their inner product with $\{v_{j} | j \in 1, 2,...,P\}$ survives. Similarly, for the set of spin orbitals $\{u_{i} | i \in P+1, P+2,...,P+Q\}$, only their inner product with $\{v_{j} | j \in P+1, P+2,...,P+Q\}$ survives. \\
The equation therefore reduces to:
\begin{equation}
\begin{aligned}
\begin{split}
	\braket { U | V } = { }& \Big( \sum_{R_{i} \in S_{P}} (-1)^{r_{i}} \braket { u_{1} | v_{R_{1}} } ... \braket { u_{P} | v_{R_{P}} } \Big) \\
	&  \Big( \sum_{R_{j} \in S_{Q}} (-1)^{r_{j}} \braket { u_{1} | v_{R_{P+1}} } ... \braket { u_{P+Q} | v_{R_{P+Q}} } \Big) \\
	= { }& det(\boldsymbol{S}_{UV}^{A})det(\boldsymbol{S}_{UV}^{B}) \\
	= { }& \braket { U^{A} | V^{A} } \braket { U^{B} | V^{B} } \\
\end{split}
\end{aligned}
\end{equation}
Since we can consider each of the sets above as permuting separately, the equation can be decomposed into a product form. Each of the terms corresponds to the overlap between states U and V centred on a particular molecule (A or B). \\

\subsection{Hamiltonian matrix}
For the Hamiltonian matrix, we separate it into the one- and two- electron cases respectively.\\
For the one-electron case:
\begin{equation}
\begin{aligned}
\begin{split}
	\braket { U | \hat{F} | V } = { }& \sum_{ij}^{P+Q} \braket { u_{i} | \hat{f} | v_{j} } D(i | j)\\
	= { }& \sum_{i,j=1}^{P} \braket { u_{i} | \hat{f} | v_{j} } \boldsymbol{S}^{A}(i | j) det(\boldsymbol{S}_{UV}^{B}) \\
	& +  \sum_{i,j=P+1}^{P+Q} \braket { u_{i} | \hat{f} | v_{j} } \boldsymbol{S}^{B}(i | j) det(\boldsymbol{S}_{UV}^{A})\\
	= { }& \braket {U^{A} | \hat{F} | V^{A} } \braket { U^{B} | V^{B} } \\
	& + \braket {U^{B} | \hat{F} | V^{B} } \braket { U^{A} | V^{A} } 
\end{split}
\end{aligned}
\end{equation}
Where we used the fact that if $u_{i}$ and $v_{j}$ belongs to different molecules, $\braket { u_{i} | \hat{f} | v_{j} } = 0$. The decomposition of $D(i|j)$ can be understood with reference to Figure \ref{fig: Matrices}. For molecule A, $1 \leq i \leq P$ and $1 \leq j \leq P$. Therefore, the bottom right block of $\boldsymbol{S}^{B}$ remains unchanged. The determinant of the whole matrix is the product of the determinant of the top left block and the bottom right block. The latter is $det(\boldsymbol{S}_{UV}^{B})$, while the former is simply the cofactor of block $\boldsymbol{S}^{A}$.\\
For the two-electron case:
\begin{equation}
\begin{aligned}
\begin{split}
	\braket { U | \hat{G} | V } = { }& \frac{1}{2}\sum_{ijkl}^{P+Q} \braket { u_{i} u_{j} | \hat{g} | v_{k} v_{l}} D(ij |kl)\\
	= { }& \sum_{ijkl=1}^{P} \braket { u_{i} u_{j} | \hat{g} | v_{k} v_{l}} \boldsymbol{S}^{A}(ij |kl) det(\boldsymbol{S}_{UV}^{B}) + \\
	&\sum_{ijkl=P+1}^{P+Q} \braket { u_{i} u_{j} | \hat{g} | v_{k} v_{l}} \boldsymbol{S}^{B}(ij |kl) det(\boldsymbol{S}_{UV}^{A}) \\
	= { }& \braket {U^{A} | \hat{G} | V^{A} }\braket { U^{B} | V^{B} } \\
	& + \braket {U^{B} | \hat{G} | V^{B} }\braket { U^{A} | V^{A} } 
\end{split}
\end{aligned}
\end{equation}
where $\braket { u_{i}(1) u_{j}(2) | \hat{g} | v_{k}(1) v_{l}(2)}$ is written as $\braket { u_{i}u_{j} | \hat{g} | v_{k} v_{l}}$ for the sake of brevity. 
Since $\hat{H} = \hat{F} + \hat{G}$,
\begin{equation}
\begin{aligned}
\begin{split}
	\braket { U | \hat{H} | V } = { }& \braket { U | \hat{F} + \hat{G} | V }\\
	= { }& \braket { U | \hat{F} | V }  + \braket { U | \hat{G} | V } \\
	= { }& \braket {U^{A} | \hat{F} + \hat{G} | V^{A} } \braket { U^{B} | V^{B} } \\
	& + \braket {U^{B} | \hat{F} + \hat{G} | V^{B} }\braket { U^{A} | V^{A} }  \\
	= { }&\braket {U^{A} | \hat{H} | V^{A} }\braket { U^{B} | V^{B} } \\
	& + \braket {U^{B} | \hat{H} | V^{B} }\braket { U^{A} | V^{A} }  
\end{split}
\end{aligned}
\end{equation}

\subsection{Forming the eigenket representation}
The construction of spin eigenfunctions is well established in literature\cite{Pauncz, Scholes, ModularTensor}. This section will provide a short summary on our approach to obtaining the eigenket representation of our SR-NOCI states. 
\subsubsection{Ladder Operators}
The formation of the eigenket representation for triplet-pair states (states which can be expressed as a linear combination of $\ket{1, a} \ket{1, b}$ where $a, b = 1, 0, -1$) is well-known. From the $\ket {1, 1} \ket {1, 1}$ state, which is one of highest spin and spin multiplicity given the coupling of two triplets, we can apply the annihilation operator, $\boldsymbol{\hat{J}_{-}}$ which is defined by:
\begin{equation}
	\boldsymbol{\hat{J}_{-}} \ket{J, M} = \sqrt{J(J+1) - M(M-1)} \ket{J, M-1}
\end{equation}
We can obtain all the nine triplet states in this manner. However, this method is not applicable for states that has a $\ket{0,0}$ component has it has a different spin. A more general method will thus be described in the next subsection.

\subsubsection{Projection onto the direct product basis}
A more general way of getting at the eigenket representation of the dimer is to construct it from that of the corresponding molecule. Denoting a spin-rotated state of the molecule as $\ket{\psi^{mol}(\theta)}$ and the spin-rotated state of the dimer as $\ket{\psi^{dimer}(\theta_{A}, \theta_{B})}$, where $\theta_{A(B)}$ represents the angle by which the spin of molecule A(B) is rotated. For the purposes of this section, the axis of rotation is assumed to be arbitrary, but kept consistent throughout.\\
From this relation,
\begin{equation}
	\ket{\psi^{dimer}(\theta_{A}, \theta_{B})} = \ket{\psi^{mol}(\theta_{A})} \otimes \ket{\psi^{mol}(\theta_{B})}
\end{equation}
we can express the spin eigenket in terms of the spin-rotated states.\\
Noting that the SR-NOCI state (these are spin purified states) for the molecule can be written as:
\begin{equation}
	\ket{\Psi^{NOCI, mol}_{i}} = \sum_{k} \ket{\psi^{mol}_{k}} ?[l]p^{k}_i?
\end{equation}
Direct products of the spin states can be found by:
\begin{equation}
	\begin{split}
	\ket { \Psi^{basis}_{i,j} } &= \ket{\Psi^{NOCI, mol}_{i}} \otimes \ket{\Psi^{NOCI, mol}_{j}}\\
	\ket { \Psi^{basis}_{m} } &= 	(\sum_{k} \ket{\psi^{mol}_{k}} ?[l]p^{k}_i?) \otimes 	(\sum_{l} \ket{\psi^{mol}_{l}} ?[l]p^{l}_j?) \\
										  &= \sum_{kl} \ket{\psi^{dimer}(\theta_{A}, \theta_{B})} (?[l]p^{k}_i? \otimes ?[l]p^{l}_j?) \\
      	                                 &= \sum_{k} \ket{\psi^{dimer}_{k}} ?[l]c^{k}_m?
	\end{split}
\end{equation}
where we have used a single index $m$ for $\ket { \Psi^{basis}}$ instead for ease of notation in the second step, and indexed $\ket{\psi^{dimer}(\theta_{A}, \theta_{B})}$ using $k$ instead of the spin rotation angles.
We can now express the SR-NOCI state, $\ket{\Psi^{NOCI, dimer}_{j}}$ as a linear combination of spin-rotated states:
\begin{equation}
	\ket{\Psi^{NOCI, dimer}_{j}} = \sum_{l} \ket{\psi^{dimer}_{l}} ?[l]d^{l}_j?
\end{equation}
To find the components of the direct product basis in the SR-NOCI state, we simply take the overlap between them:
\begin{equation}
	\braket { \Psi^{basis}_{m} | \Psi^{NOCI, dimer}_{j}} = \sum_{k,l} ?[l] c*_{m}^{k}? \braket { \psi^{dimer}_{k} | \psi^{dimer}_{l} } ?[l] d^{l}_{j}?
\end{equation}
This will project out the components of the SR-NOCI state which can inform us of the spin eigenket representation of each SR-NOCI state. It should be stressed that this approach only works when the two molecules in the dimeric system are well-separated insofar as the overlap between the two molecules approaches zero. 

\bibliography{Bibliography}

\section{For Table of Contents Only}
\includegraphics[width=8.25cm, height=4.45cm]{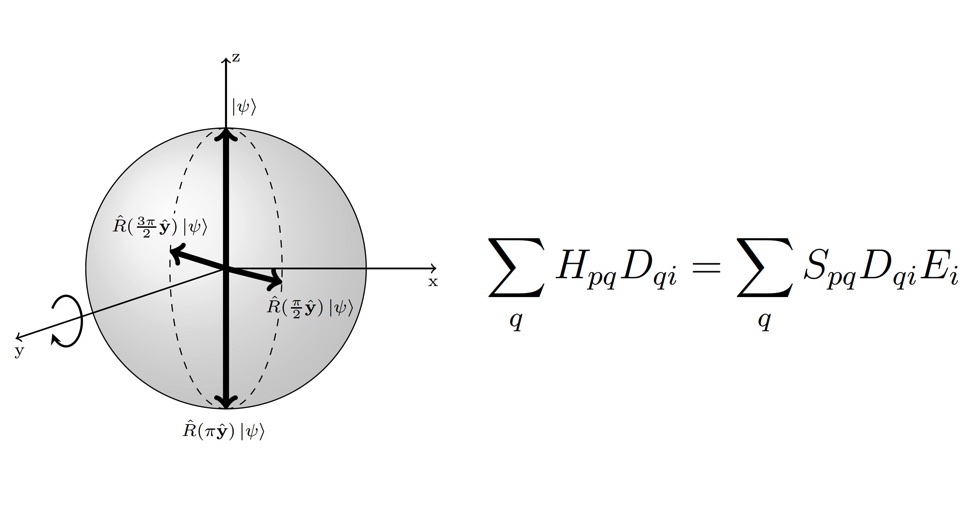}
\end{document}